\documentclass[pdftex,twocolumn,epjc3,final]{svjour3}  

\pdfoutput=1 

\usepackage{preprintcover} 
\PreprintCoverPaperTitle{Measurement of flow harmonics with multi-particle cumulants in Pb+Pb collisions at $\sqrt{s_{\mathrm{NN}}}=2.76$~\TeV\  with the ATLAS detector} 
 \PreprintIdNumber{CERN-PH-EP-2014-171} 
 \PreprintCoverAbstract{ATLAS measurements of the azimuthal anisotropy in lead--lead collisions at $\sqrt{s_{\mathrm{NN}}}=2.76$ \TeV\ are shown using a dataset of approximately 7 $\mu$b$^{-1}$ collected at the LHC in 2010.  The measurements are performed for charged particles with transverse momenta $0.5<p_{\mathrm{T}}<20$~\gev\ and in the pseudorapidity range $|\eta|<2.5$. The anisotropy is characterized by the Fourier coefficients, $\mathrm{v}_n$, of the charged-particle azimuthal angle distribution for $n$ = 2--4.  The Fourier coefficients are evaluated using multi-particle cumulants calculated with the generating function method. Results on the transverse momentum, pseudorapidity and centrality dependence of the $\mathrm{v}_n$ coefficients are presented. The elliptic flow, $\mathrm{v}_2$, is obtained from the two-, four-, six- and eight-particle cumulants while higher-order coefficients, $\mathrm{v}_3$ and $\mathrm{v}_4$, are determined with two- and four-particle cumulants.   Flow harmonics $\mathrm{v}_n$ measured with  four-particle cumulants are significantly reduced compared to the measurement involving two-particle cumulants. A comparison to $\mathrm{v}_n$ measurements obtained using different analysis methods and previously reported by the LHC experiments is also shown.  
Results of measurements of flow fluctuations  evaluated with multi-particle cumulants are shown as a function of transverse momentum and the collision centrality. Models of the initial spatial geometry and its fluctuations fail to describe the flow fluctuations measurements.
 } 
\PreprintJournalName{EPJ C}

\renewcommand\figurename{Fig.}

\RequirePackage{graphicx}

\usepackage[numbers,sort&compress]{natbib}
\usepackage{hyperref}
\usepackage{mathptmx}
\usepackage{graphicx}
\usepackage{atlasphysics}
\usepackage[switch]{lineno}
\usepackage{cite}

\journalname{Eur. Phys. J. C}

\begin{document}
\setlength{\emergencystretch}{4em} 
\setlength{\columnwidth}{75mm}

\title{Measurement of flow harmonics with multi-particle cumulants in Pb+Pb collisions at $\sqrt{s_{\mathrm{NN}}}=2.76$~\TeV\  with the ATLAS detector}

\author{The ATLAS Collaboration}
\institute{The ATLAS Collaboration \at CERN, 1211 Geneva 23, Switzerland}

\date{Received: 19.8.2014 }

\maketitle

\begin{abstract}
ATLAS measurements of the azimuthal anisotropy in lead--lead collisions at $\sqrt{s_{\mathrm{NN}}}=2.76$ \TeV\ are shown using a dataset of approximately 7 $\mu$b$^{-1}$ collected at the LHC in 2010.  The measurements are performed for charged particles with transverse momenta $0.5<p_{\mathrm{T}}<20$~\gev\ and in the pseudorapidity range $|\eta|<2.5$. The anisotropy is characterized by the Fourier coefficients, $\mathrm{v}_n$, of the charged-particle azimuthal angle distribution for $n$ = 2--4.  The Fourier coefficients are evaluated using multi-particle cumulants calculated with the generating function method. Results on the transverse momentum, pseudorapidity and centrality dependence of the $\mathrm{v}_n$ coefficients are presented. The elliptic flow, $\mathrm{v}_2$, is obtained from the two-, four-, six- and eight-particle cumulants while higher-order coefficients, $\mathrm{v}_3$ and $\mathrm{v}_4$, are determined with two- and four-particle cumulants.   Flow harmonics $\mathrm{v}_n$ measured with  four-particle cumulants are significantly reduced compared to the measurement involving two-particle cumulants. A comparison to $\mathrm{v}_n$ measurements obtained using different analysis methods and previously reported by the LHC experiments is also shown.  
Results of measurements of flow fluctuations  evaluated with multi-particle cumulants are shown as a function of transverse momentum and the collision centrality. Models of the initial spatial geometry and its fluctuations fail to describe the flow fluctuations measurements.
\end{abstract}

\section{Introduction}
\label{sec:intro}
The anisotropy of charged-particle azimuthal angle distributions in heavy-ion collisions has been a subject of extensive experimental studies at RHIC  \citep{flowReview,flowReview1,brahms1,phobos1,phenix1,star1} and more recently at the LHC  \citep{Alice1,Alice2,Alice3,Alice4,Alice5,Alice6,Alice7,Alice8,Atlas1,Atlas2,Atlas3,Atlas4,Atlas5,Cms1,Cms2,
Cms3,Cms4,Cms5}. The results provide conclusive evidence that the hot and dense matter produced in these collisions behaves collectively and has properties resembling those of a nearly perfect fluid~\citep{HeinzRev}.

The final-state anisotropy is a global property of particle production that arises from the initial spatial asymmetry of the collision region in a plane transverse to the beam axis for heavy-ion collisions with a non-zero impact parameter. It is characterized by the coefficients, $\mathrm{v}_n$, of the Fourier expansion of the measured azimuthal angle distributions \citep{flowReview,Poskanzer:1998yz}:
\begin{equation}
\mathrm{v}_n \equiv \langle {\rm e}^{\mathrm{i}n(\phi - \Psi_n)}\rangle = \langle \cos{[n(\phi - \Psi_n)]}\rangle,
\label{eq:vn}
\end{equation}
\noindent
where $n$ is the order of the Fourier harmonic, referred to as flow harmonic, $\phi$ is the azimuthal angle of the outgoing particle, $\Psi_n$ defines the azimuthal angle of the $n$th-order symmetry plane of the initial geometry~\citep{Atlas1}, and the angled brackets denote an average over charged particles in an event. Due to the symmetry in the azimuth of the plane defined by $\Psi_n$, all sine terms of the Fourier expansion vanish.  For evaluation of the coefficients $\mathrm{v}_{n}$ in the ``event-plane" method, the initial plane of symmetry is estimated from the measured correlations between particles, using the so-called sub-event method~\citep{Poskanzer:1998yz}. As a consequence, only the two-particle correlations are exploited in the determination of $\mathrm{v}_{n}$ (see Eq.~(\ref{eq:vn})). This leads to a problem of disentangling all-particle flow and contributions from particle correlations unrelated to the initial geometry, known as non-flow correlations. These non-flow effects include correlations due to energy and momentum conservation, resonance decays, quantum interference phenomena and jet production. They generally involve only a small number of produced particles. In order to suppress non-flow correlations, methods that use genuine multi-particle correlations, estimated using cumulants, were proposed~\citep{Borghini:2000,Borghini:2001,Borghini:guide,Snellings}. 

Calculating multi-particle correlations in large-multiplicity heavy-ion collisions at high energies is limited by the computing requirements needed to perform nested loops over thousand of particles per event to analyse all particle multiplets. 
To avoid this problem, the generating function formalism~\citep{Borghini:2000,Borghini:2001,Borghini:guide} is exploited to calculate multi-particle cumulants, and the results obtained are presented in this paper. An alternative approach was proposed in Ref.~\citep{Snellings} to express multi-particle correlations in terms of the moments of the flow vector, $Q_n$, and is used in this paper as a cross-check of multi-particle cumulants obtained with the generating function method.
The cumulant approach to measure flow harmonics also provides the possibility to study  event-to-event fluctuations in the amplitudes of different harmonics, which  can be related to the fluctuations in the initial transverse shape of the interaction region~\citep{phobosfluct1,phobosfluct2,starfluct}. 

The cumulant method has been used to measure the anisotropic flow in NA49 \citep{spscum}, STAR \citep{rhiccum} and recently also at the LHC experiments~\citep{Alice1,Alice3,Cms1,Cms4}. The results show that the Fourier coefficients determined with four-particle cumulants are smaller than those derived with two-particle cumulants due to the suppression in the former of non-flow two-particle correlations. In this paper, the  method is used to measure flow harmonics in lead--lead collisions at $\sqrt{s_{\mathrm{NN}}}=2.76$~\TeV\ with the ATLAS detector. The elliptic flow $\mathrm{v}_2$ is measured using two-, four-, six- and eight-particle cumulants. For $\mathrm{v}_3$ and $\mathrm{v}_4$ measurements the two- and four-particle cumulants are exploited.

This paper is organized as follows. Section~\ref{sec:samples} describes the ATLAS detector, trigger, and offline event selections. Section~\ref{sec:eventsTracks} contains a description of additional selection criteria for events and charged-particle tracks.  Section~\ref{sec:MC} gives details of the Monte Carlo simulation samples used to derive the tracking efficiency and fake-track rates.  The analysis method and procedure is outlined in Sect.~\ref{sec:analysis}. Section~\ref{sec:systematics} contains a discussion of the systematic errors. Results are presented in Sect.~\ref{sec:results}. 
Section~\ref{sec:summary} is devoted to summary and conclusions.

\section{The ATLAS detector and trigger}
\label{sec:samples}
The results presented in this paper were obtained from a sample of minimum-bias lead--lead collisions at $\sqrt{s_{\mathrm{NN}}}=2.76$~\TeV\ recorded by ATLAS in 2010 and corresponding to an integrated luminosity of approximately $7$~$\mu\mathrm{b}^{-1}$. The measurements were performed using the ATLAS inner detector and forward calorimeters~\citep{Aad:2008zzm}. The inner detector covers the complete azimuthal range and extends over the pseudorapidity region $|\eta|<2.5$.\footnote{ATLAS uses a right-handed coordinate system with its origin at the nominal interaction point (IP) in the centre of the detector and the $z$-axis along the beam pipe. The $x$-axis points from the IP to the centre of the LHC ring, and the $y$-axis points upward. Cylindrical coordinates $(r,\phi)$ are used in the transverse plane, $\phi$ being the azimuthal angle around the beam pipe. The pseudorapidity is defined in terms of the polar angle $\theta$ as $\eta=-\ln\tan(\theta/2)$.} The inner detector silicon tracker, used in this analysis for track reconstruction, consists of layers of pixel and microstrip detectors (SCT) immersed in a $2\,$T axial magnetic field. The forward calorimeters (FCal) use liquid argon with copper-tungsten absorbers to perform both the electromagnetic and hadronic energy measurements with copper-tungsten/liquid argon technology, and also provide complete coverage in azimuth for $3.2<|\eta|<4.9$. The trigger system was used to select minimum-bias lead--lead collisions. It required a coincidence of signals recorded in both zero-degree calorimeters (ZDC), located symmetrically at $z=\pm140$~m, and in the minimum-bias trigger scintillator (MBTS) counters at $z=\pm3.6$~m.

\section{Event and track selections}
\label{sec:eventsTracks}
Additional offline event selections were also applied, requiring a time difference between the two MBTS counters of less than $3$~ns and at least one primary vertex reconstructed using charged-particle tracks. 
 Events satisfying the above-described selections were also required to have a reconstructed primary vertex within $100$~mm of the nominal centre of the ATLAS detector. 

The precision silicon tracking detectors  were used to reconstruct charged-particle trajectories with a minimum \pT\ of $0.5\,$\gev. Special track-quality criteria are imposed to deal with high particle densities in Pb+Pb collisions. 
Tracks are required to have at least eight hits in the SCT, at least two pixel hits and a hit in the pixel layer closest to the interaction point if expected. A track must have no missing pixel hits and no missing SCT hits, when a hit is expected. The transverse and longitudinal impact parameters with respect to the vertex, $|d_0|$ and $|z_0\sin\theta|$ respectively, were each required to be less than 1 mm. Specifically for this analysis it was also required that $|d_0/\sigma_{d_0}|<3$ and $|z_0\sin{\theta}/\sigma_z|<3$, where $\sigma_{d_0}$ and $\sigma_z$ are the uncertainties on $d_0$ and $z_0\sin\theta$, respectively, as obtained from the covariance matrix of the track fit. The latter requirements improve both the tracking performance at high $p_{\mathrm{T}}$ and the purity of the track sample. The number of reconstructed tracks per event is denoted $N_{\mathrm{ch}}^{\mathrm{rec}}$. For this analysis, the additional requirement of $ N_{\mathrm{ch}}^{\mathrm{rec}} \geq 10$ for tracks with $0.5<p_{\mathrm{T}}<5\,$\gev\ was imposed to allow the measurement of correlations involving as many as eight particles. 

The correlation between the summed transverse energy ({\mbox{$\Sigma E_{\mathrm{T}}^{\mathrm{FCal}}$}}) measured in the FCal and $N_{\mathrm{ch}}^{\mathrm{rec}}$ was investigated in order to identify background events. Events having an $N_{\mathrm{ch}}^{\mathrm{rec}}$ vs. {\mbox{$\Sigma E_{\mathrm{T}}^{\mathrm{FCal}}$}} correlation distinctly different from that for the majority of Pb+Pb collisions were removed. The removed events, less than 0.01\% of the sample, were found to contain multiple Pb+Pb collisions.  After applying all selection requirements, the data sample consists of about $35 \times 10^{6}$ Pb+Pb collision events. 

The summed transverse energy is used to define the centrality of the collision. A detailed analysis of the {\mbox{$\Sigma E_{\mathrm{T}}^{\mathrm{FCal}}$}} distribution \citep{Atlas1} showed that the fraction of the total inelastic cross-section sampled by the trigger and event selection requirements is $(98 \pm 2)$\%.  The {\mbox{$\Sigma E_{\mathrm{T}}^{\mathrm{FCal}}$}} distribution was divided into centrality intervals, each representing a percentile fraction of all events after accounting for the 2\% inefficiency in recording the most peripheral collisions. The analysis is performed in narrow centrality intervals: 1\% centrality bins for the 20\% of events with the largest {\mbox{$\Sigma E_{\mathrm{T}}^{\mathrm{FCal}}$}}, and 5\% centrality bins for the remaining events. These narrow centrality intervals are then combined into wider bins to ensure sufficiently small statistical uncertainties on the measured flow harmonics. The 20\% of events with the smallest {\mbox{$\Sigma E_{\mathrm{T}}^{\mathrm{FCal}}$}} (most peripheral collisions) are not considered in this analysis, due to the  inefficiency in the event triggering and the correspondingly large uncertainties of measurements performed for these low-multiplicity collisions. 
For each centrality interval, a standard Glauber Monte Carlo model \citep{glauber1,tracklets} is used to estimate the average number of participating nucleons, $\langle N_{\mathrm{part}}\rangle$, which provides an alternative measure of the collision centrality.

\section{Monte Carlo simulations}
\label{sec:MC}
A Monte Carlo (MC) sample was used in the analysis to determine tracking efficiencies and rates of falsely reconstructed  tracks (fake-track rates). The HIJING event generator \citep{Gyulassy:1994ew} was used to produce minimum-bias Pb+Pb collisions. Events were generated with the default parameters, except for the jet quenching, which was turned off. 
 Flow harmonics were introduced into HIJING at the generator
level by changing the azimuthal angle of each particle \citep{flowReview} in order to produce an anisotropic azimuthal angle distribution consistent with previous ATLAS $\mathrm{v}_n$ ($n$ = 2--6) measurements~\citep{Atlas1,Atlas2}. The detector response simulation~\citep{AtlasSim}  uses the GEANT4 package \citep{GEANT4} with data-taking conditions corresponding to those of the 2010 Pb+Pb run and simulated events are reconstructed in the same way as the data.

The tracking efficiency, $\epsilon(\pT ,\eta)$, and the fake-track rate $f(\pT ,\eta)$ are determined \citep{trackingATLAS} using the Monte Carlo sample described above. The MC reproduces the measured centrality dependence of the track-quality parameters. The efficiency is found to depend weakly on the collision centrality. For the lowest transverse momenta (0.5--0.6~\gev), the efficiency at $|\eta|<1$ is of the order of 50\% and falls to about 30\% at high $|\eta|$. For higher transverse momenta it reaches about 70\% at $|\eta|<1$ and drops to about 50\% at high $|\eta|$. The rate of falsely reconstructed tracks (the fake-track rate) is typically below 1\%.  It increases to 3--7~\% for the lowest transverse momenta in the most central collisions.

\section{Analysis procedure}
\label{sec:analysis}
Fourier coefficients, $\mathrm{v}_n$, are measured using $2k$-particle correlations \citep{Borghini:2000,Borghini:2001,Borghini:guide} defined as:
\begin{eqnarray}
\langle corr_n\{2k\}\rangle & = &\langle\langle {\rm e}^{\mathrm{i}n(\phi_1+\ldots+\phi_k-\phi_{1+k}-\ldots-\phi_{2k})}\rangle\rangle \nonumber\\
&=&\langle \mathrm{v}_n\{2k\}^{2k} \rangle,
\label{eq:corr2k}
\end{eqnarray}
\noindent
where the notation $\mathrm{v}_n\{2k\}$ is used for the $\mathrm{v}_n$ flow harmonic derived from  the $2k$-particle correlations, and $k$ is an integer. Azimuthal angles of particles forming a $2k$-particle cluster are denoted by $\phi_l$, where $l=1, \dots, 2k$. The double angled brackets denote an average, first over charged particles in an event, and then over events, while the single angled brackets denote averaging over events.
 The multi-particle correlation, $\langle corr_n\{2k\}\rangle$, includes contributions from the collective anisotropic flow and from non-flow effects (see Sect.~\ref{sec:intro}).  It was proposed in Refs.~\citep{Borghini:2000,Borghini:2001,Borghini:guide} to exploit the cumulant expansion of multi-particle correlations in order to reduce the non-flow contribution. The anisotropic flow related to the initial geometry is a global, collective effect involving correlations between all outgoing particles. Thus, in the absence of non-flow effects, $\mathrm{v}_n\{2k\}$ is expected to be independent of $k$. On the other hand, most of the non-flow correlations, such as resonance decays or interference effects, contribute only to correlations between small numbers of particles. The idea of using $2k$-particle cumulants is to suppress the non-flow contribution by eliminating the correlations which act between fewer than $2k$ particles. More specifically, the cumulant of e.g. the four-particle correlations, defined as:
\begin{equation}
c_n\{4\}=\langle corr_n\{4\}\rangle -2\langle corr_n\{2\}\rangle^2,
\label{eg:c4}
\end{equation}
\noindent
measures the genuine four-particle correlations. So, if the non-flow contribution is only due to the two-particle correlations, then $c_n\{4\}$ directly measures flow harmonics.  Similarly, using the cumulant of the six-particle correlations allows one to remove contributions from two- and four-particle correlations. The different cumulants provide independent estimates of the same flow harmonic $\mathrm{v}_n$, with the estimate based on correlations among many particles being more precise due to the suppressed non-flow correlations. In the absence of non-flow correlations, cumulants of different order should give the same estimate of $\mathrm{v}_n$.

The generating function formalism for calculating $2k$-particle cumulants (GFC method) was proposed in Ref.~\citep{Borghini:guide}. With this method, the number of required computing operations is proportional to the number of particles per event. The cumulant generating function of multi-particle azimuthal correlations, $C_n(z)$, is defined in the plane of a complex variable $z$ as:
\clearpage
\begin{eqnarray}
 \lefteqn {C_n(z)  =<N>} \nonumber \\
  & & \times \left ( \left<\prod_{j=1}^{N}\left[1 + \frac{w_j(z{\rm e}^{\mathrm{i}n\phi_j}+z^*{\rm e}^{-\mathrm{i}n\phi_j})}{N}\right]\right>^{1/<N>} -1 \right),
  \label{eq:gfc}
\end{eqnarray}
where the angled brackets represent the average over events in a given centrality interval, and the product runs over the $N$ particles within a given Pb+Pb event \citep{Borghini:2000,Borghini:2001,Borghini:guide}. The weighting factors, $w_j$, are used in this analysis to correct for any non-uniformity in the azimuthal angle distribution of reconstructed tracks. The weights are obtained from the data using the two-dimensional distribution in the $\eta$--$\phi$  plane of all reconstructed tracks. For each bin $j$ in $(\delta\eta,\delta\phi)=(0.1,2\pi/64)$ a weight is calculated as $w_j=\left\langle N(\delta\eta)\right\rangle /N(\delta\eta,\delta\phi)$, where $\left\langle N(\delta\eta)\right\rangle $ is the average number of tracks in the $\delta\eta$ slice to which this bin belongs, while $N(\delta\eta,\delta\phi)$ is the number of tracks in the $(\delta\eta,\delta\phi)$ bin. 

The expansion of the cumulant generating function in powers of $|z|$ provides the cumulant $c_n\{2k\}$, which is equal to the coefficient of the term $|z|^{2k}/k!^2$ of this expansion. In practice, to construct the $c_n\{2k\}$ cumulant the power series is truncated to order $|z|^{2k}$ and $C_n(z)$ is computed at a discrete set of interpolating points $z_{p,q}=x_{p,q}+iy_{p,q}$ \citep{Borghini:guide}, where:
\begin{eqnarray}
  x_{p,q}=r_0\sqrt{p}\cdot\cos\left(\frac{2q\pi}{q_{\rm{max}}}\right),  \\ y_{p,q}=r_0\sqrt{p}\cdot\sin\left(\frac{2q\pi}{q_{\rm{max}}}\right).
  \label{eq:points}
\end{eqnarray}
\noindent
For this analysis, the parameters $p=1,...,5$ and $q=0,...,q_{\rm{max}}-1$ with $q_{\rm{max}}=11$ were chosen as recommended in Ref.~\citep{Borghini:guide}. The $r_0$ parameter ($r_0\equiv |z|/\sqrt{p}$) should be as small as possible, chosen such that the results remain stable under its variation.  The $r_0$ values used were chosen to be 4.0, 2.2, 1.6, 1.1 and 1.0 for centrality intervals 0--5\%, 5--10\%, 10--20\%, 20--30\% and 30--80\%, respectively. For these values, the cumulants are found to be stable when varying $r_0$ between $r_0/2$ and $2r_0$. The only differences, up to about 2\%, were seen when using the eight-particle cumulants to calculate the elliptic flow harmonic and are accounted for in the systematic uncertainty on $\mathrm{v}_2\{8\}$.

An alternative method to calculate multi-particle correlations and cumulants in a single pass over all particles in each event, referred to as the QC method, was proposed in Ref.~\citep{Snellings}. In this method, the expressions for the multi-particle correlations are derived in terms of the moments of the flow vector $Q_n$, defined as $Q_n=\sum_{j=1}^{N} w_j {\rm e}^{\mathrm{i}n\phi_j}$, where the index $n$ denotes the order of the flow harmonic, the sum runs over all $N$ particles in an event and $w_j$ are weights as defined above. The QC method is used to calculate the cumulants, $c_n\{2k\}$, which are compared with the cumulants obtained from the GFC method.
\begin{figure*}[th!]
\begin{center}
\includegraphics[width=120mm]{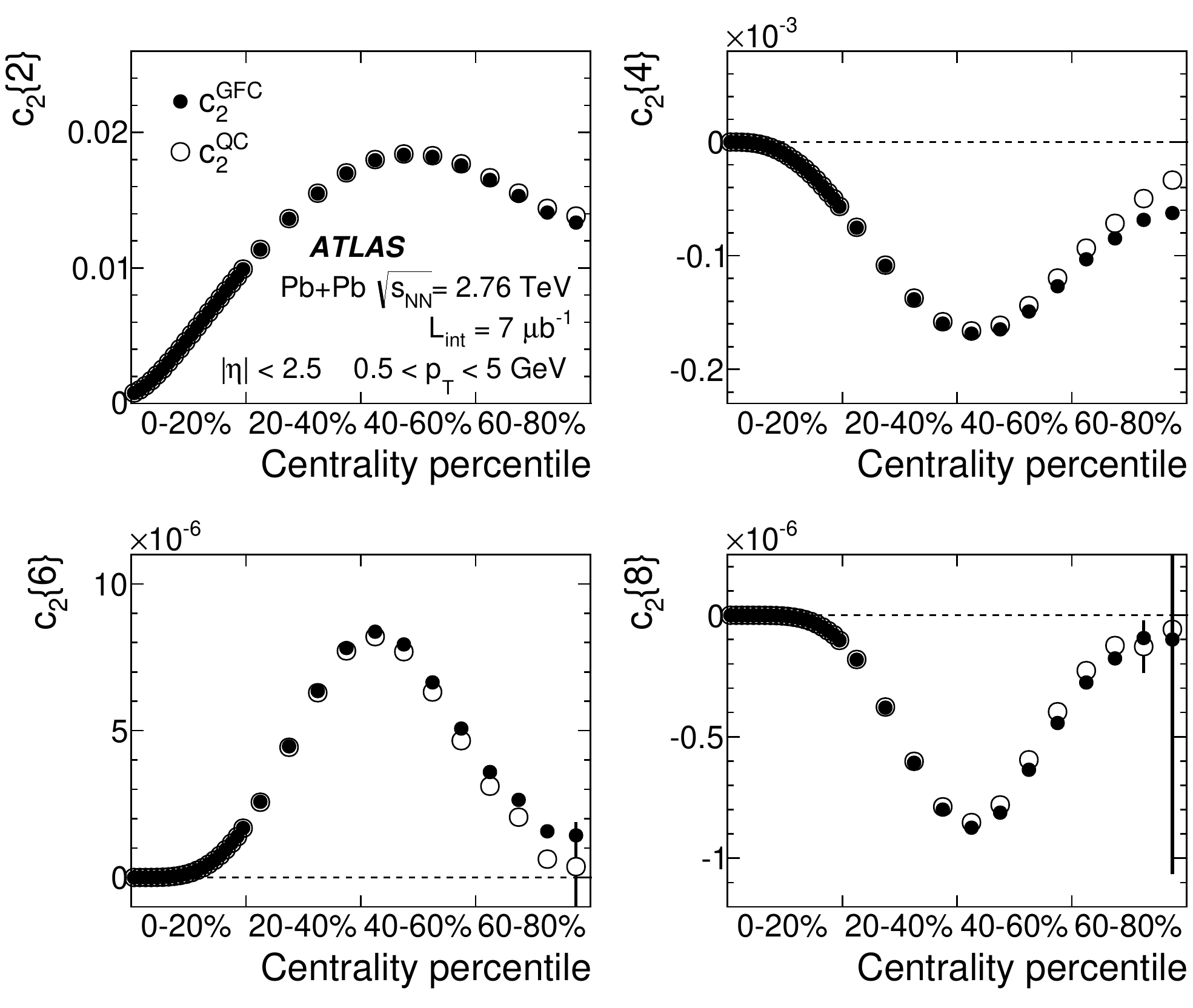}
\caption{Multi-particle cumulants for the second-order flow harmonic, $ c_2\{2k\}$ for $k=1,2,3,4$, obtained with the GFC \citep{Borghini:guide} and QC \citep{Snellings} methods shown as a function of centrality.  The horizontal axis ranges from central collisions (0--20\%) to more peripheral collisions (60--80\%).}
\label{fig:Fig2} 
\end{center}
\end{figure*}

A practical application of the cumulant method involves two main steps \citep{Borghini:2000,Borghini:2001,Borghini:guide}. First, the reference  $2k$-particle cumulants, $c_n\{2k\}$, are derived from the cumulant generating function calculated from particles measured over a broad range of transverse momentum and pseudorapidity.
This step is equivalent to the event-plane estimate in the standard method (see Eq.~(\ref{eq:vn})) and the reference cumulants play a similar role to the event-plane resolution correction~\citep{Poskanzer:1998yz}. 
In the next step, the differential flow is calculated in \pT\ and $\eta$ bins using cumulants, denoted $d_n\{2k\}$, computed from a differential generating function. To determine the $d_n\{2k\}$ cumulants, each charged particle from a \pT\ and $\eta$ bin is correlated with $2k-1$ reference particles. The differential flow harmonics $\mathrm{v}_n\{2k\}(\pT,\eta)$, are then calculated with respect to the reference cumulants as prescribed in Refs.~\citep{Borghini:2001,Borghini:guide}:
 \begin{eqnarray}
&&  \label{eq:dvn2} \mathrm{v}_n\{2\}(\pT,\eta)=\frac{d_n\{2\}}{\sqrt{c_n\{2\}}},\\
&&  \label{eq:dvn4} \mathrm{v}_n\{4\}(\pT,\eta)=\frac{-d_n\{4\}}{\sqrt[3/4]{-c_n\{4\}}},\\
&&  \label{eq:dvn6} \mathrm{v}_n\{6\}(\pT,\eta)=\frac{d_n\{6\}/4}{\sqrt[5/6]{c_n\{6\}/4}},\\
&&  \label{eq:dvn8} \mathrm{v}_n\{8\}(\pT,\eta)=\frac{-d_n\{8\}/33}{\sqrt[7/8]{-c_n\{8\}/33}}.
  \end{eqnarray}

In order to calculate the reference cumulants, $c_n\{2k\}$, all charged particles with pseudorapidities  $|\eta|<2.5$ and transverse momenta $0.5 < \pT < 5\,$\gev\ are used in this analysis. The results for $c_2$ are shown as a function of centrality in Fig.~\ref{fig:Fig2} for two-, four-, six- and eight-particle cumulants obtained from the GFC and QC methods. The figure shows that the two methods yield consistent results over a wide range of collision centralities.  Differences, up to $\sim$ 20\%,  are observed only for the most peripheral collisions. For the most central (0--2\%) Pb+Pb collisions, the cumulants $c_n\{2k\}$ for $k>1$ are, within sizeable statistical errors, consistent with zero. However,  they have incorrect signs, which prevents the calculation of flow harmonics due to the square-root function in the denominator of Eqs.~(\ref{eq:dvn4}), (\ref{eq:dvn6}) and (\ref{eq:dvn8}).

\begin{figure*}[ht!]
\begin{center}
\includegraphics[width=120mm]{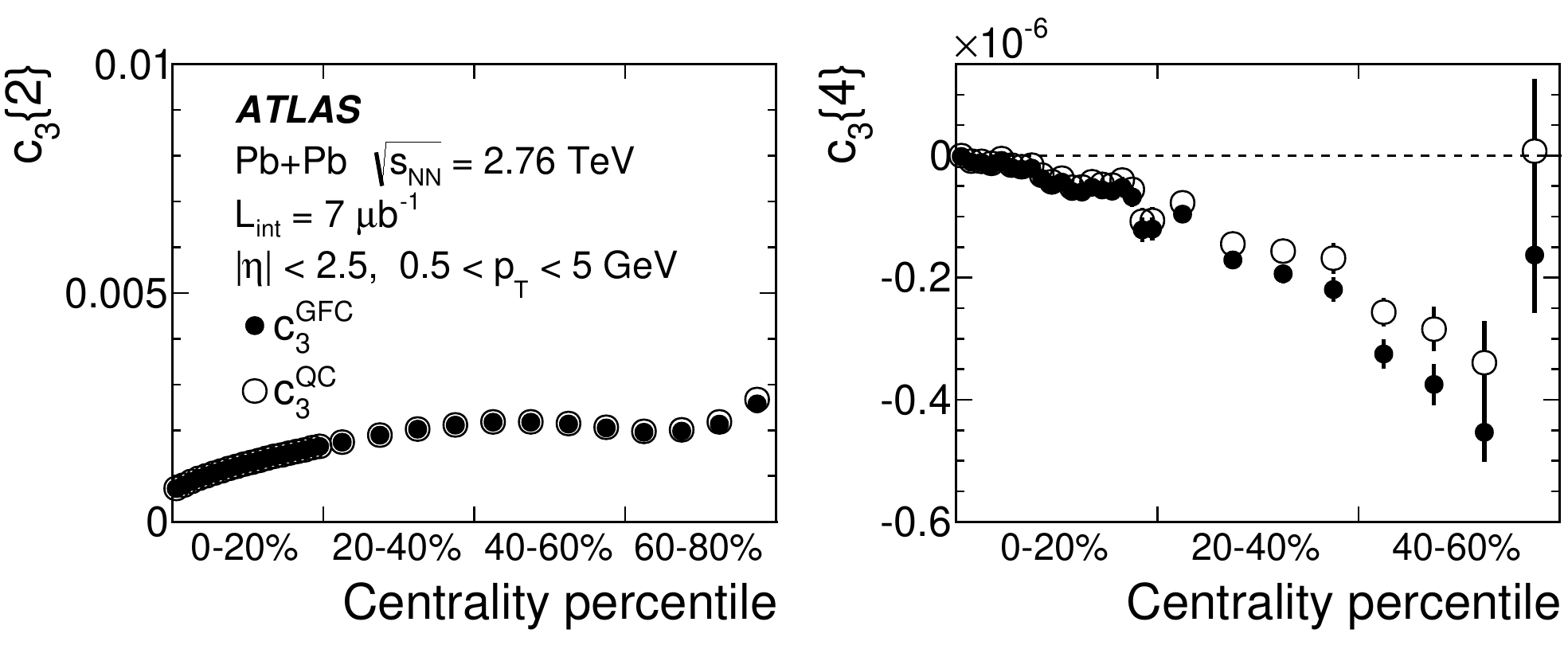}
\includegraphics[width=120mm]{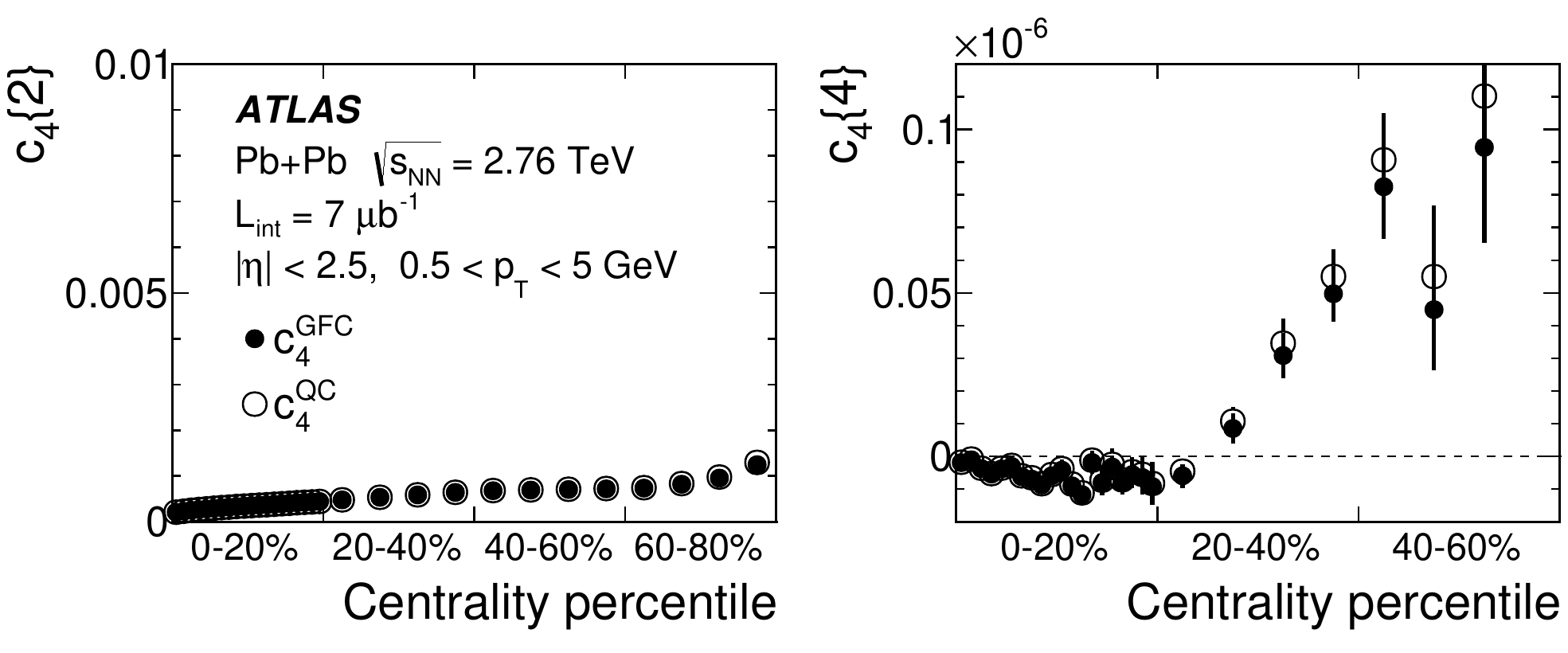}
\caption{Top: multi-particle cumulants for the third-order flow harmonic, $ c_3\{2k\}$ for $k=1,2$ obtained with the GFC \citep{Borghini:guide} and QC \citep{Snellings} methods shown as a function of centrality. Bottom: the same for the fourth-order flow harmonics, $ c_4\{2k\}$ .}
\label{fig:Fig3} 
\end{center}
\end{figure*}

For higher flow harmonics, the cumulants $c_3\{2k\} $ and $c_4\{2k\} $ obtained from both the GFC and QC methods are consistent with zero for $k>2$. Therefore, only two- and four-particle cumulants can be used to derive third- and fourth-order flow coefficients. Figure~\ref{fig:Fig3} shows the centrality dependence of the two- and four-particle cumulants, obtained from the GFC and QC methods, for $n=3$ and $n=4$. The figure demonstrates an overall good agreement between the cumulants calculated using the two different methods. In the case of four-particle cumulants, the centrality range of the method's applicability is limited to 0--60\% for $n=3$ and 0--25\% for $n=4$. 

The differential flow harmonics, $\mathrm{v}_n\{2k\}(\pT,\eta)$, are determined using the differential cumulants $d_n\{2k\}$ and Eqs.~(\ref{eq:dvn2})--(\ref{eq:dvn8}) in bins of transverse momentum and pseudorapidity for events from a given centrality interval. The pseudorapidity range $|\eta|<2.5$ is divided into 50 bins of width 0.1 each. In transverse momentum, 28 bins of variable width,  covering the \pT\ range from 0.5~\gev\ to 20~\gev, are used. These differential flow harmonics can then be integrated over wider phase-space bins or the full range in either pseudorapidity or transverse momentum, or both. In this integration procedure, the harmonics $\mathrm{v}_n\{2k\}(\pT,\eta)$ measured in each small bin are weighted by the charged-particle multiplicity in that bin, corrected for tracking efficiency and fake-track rate, using the MC-determined corrections $\epsilon(\pT ,\eta)$ and $f(\pT ,\eta)$ as described in Sect.~\ref{sec:MC}. 
\section{Systematic uncertainties}
\label{sec:systematics}
The systematic uncertainties on the measurements presented in this paper are evaluated by varying different aspects of the analysis and comparing the results obtained to the baseline results for the transverse momentum, pseudorapidity and centrality dependence of  $\mathrm{v}_2$, $\mathrm{v}_3$ and $\mathrm{v}_4$. The following sources are considered as potential contributors to the systematic uncertainty on the measured flow harmonics. 

An overall scale uncertainty on flow harmonics comes from the uncertainty in the fraction of the total inelastic cross-section accepted by the trigger and the event selection criteria. It is evaluated by varying the centrality bin definitions, using the modified selections on {\mbox{$\Sigma E_{\mathrm{T}}^{\mathrm{FCal}}$}}, which account for the 2\% uncertainty in the sampled fraction of the cross-section. 

All formulas applied in the analysis are valid under the assumption that the sine terms in the Fourier expansion vanish due to the azimuthal symmetry of the initial geometry. However, due to some distortions in the detector acceptance and non-uniformities in the measured azimuthal angle distributions, a residual sine term may be present. The magnitude of the sine term is calculated as the imaginary part of the differential generating function. The deviation from zero of the average sine term with respect to $\langle \mathrm{v}_n \rangle$ is treated as the systematic uncertainty. Some detector distortions may lead to an asymmetry between positive and negative $\eta$ hemispheres, and the difference between  the flow harmonics measured at positive and negative pseudorapidities is also considered as a systematic uncertainty. 

A small contribution to the systematic uncertainty, only for $\mathrm{v}_2\{8\}$, comes from the stability of the results with respect to the assumed value of the $r_0$ parameter (see discussion in the previous section). The correction applied to ensure the uniformity of the azimuthal angle distribution of reconstructed tracks (via weights $w_j$) is also checked by comparing the baseline results to those obtained with $w_j\equiv 1$. The contribution to the systematic uncertainty related to the track-quality definition is evaluated by comparing results obtained with more restrictive or less restrictive requirements. Both the transverse and longitudinal impact parameter cuts, $|d_0|$ and $|z_0 \sin{\theta}|$, are changed by $\pm 0.5$~mm with respect to the nominal value of  1~mm and the significance cuts, $|d_0/\sigma_{d_0}|<3$ and $|z_0\sin{\theta}/\sigma_z|<3$,  are changed by $\pm 1$. 

The analysis procedure is also checked through MC studies by comparing the observables at the generator/particle level with those obtained in the MC simulated sample for which the same analysis chain and correction procedure is used as for the data. The measured flow harmonics in data agree qualitatively with the reconstructed MC harmonics and show similar trends as a function of $\eta$. In the phase-space region where tracking performance suffers from low efficiency and high fake-track rates ($\pT < 1.5$~GeV and  
$|\eta| > 1$), systematic differences are observed between the flow harmonics calculated at the generator level and at the reconstruction level after the corrections. In general, in this 
phase-space region, the reconstructed flow harmonics are smaller than the generator-level ones and show an $\eta$ dependence, not present at the generator level. To account for this $\eta$-dependent bias, the MC closure test is considered as part of the systematic uncertainty. 

A significant systematic uncertainty on the transverse momentum dependence of $\mathrm{v_2}$ due to the centrality bin definitions was found. In the most peripheral 60--80\% centrality interval it is of the order of 5\%  for $\mathrm{v}_2\{4\}$ and rises to 14\% for $\mathrm{v}_2\{8\}$. For the most central collisions the uncertainty is in the range 1-2\%.   At low \pT, below 1.5~\gev, the systematic uncertainty due to the Monte Carlo closure is significant in the most central collisions, and reaches 8\% for $\mathrm{v}_2\{2\}$. For $\mathrm{v}_2\{2k\}$ with $k>1$ it is at the level of 3--4\%. The MC closure at \pT\ above 1.5~ \gev\ gives 4\% for the most central collisions, and stays typically below 1\% for other collision centralities. The $r_0$ stability adds about 2\% uncertainty only for $\mathrm{v}_2\{8\}$. All other considered sources give contributions well below 1\% to the systematic uncertainty on the \pT\ dependence of $\mathrm{v}_2$. For higher-order flow harmonics, the systematic uncertainty on the transverse momentum dependence is mainly due to the non-zero sine term and the MC closure. The former contributes up to 5\% (15\%) for $\mathrm{v}_3\{4\}$ ($\mathrm{v}_4\{4\}$) and about 1\% for $\mathrm{v}_3\{2\}$ and $\mathrm{v}_4\{2\}$. The uncertainty due to the MC closure is less than 6\% for $\mathrm{v}_3$, and increases to 13\% for $\mathrm{v}_4$. Contributions from other sources are of the order of 1--2\%.

The systematic uncertainty on the pseudorapidity dependence of $\mathrm{v}_2$ is dominated by the MC closure at $|\eta|>1$ (up to 7\% for $\mathrm{v}_2\{2\}$ in the most central collisions). For $\mathrm{v}_2$ calculated with six- and eight-particle cumulants, significant contributions come also from the sine term (up to 15\%), $\eta$ asymmetry (up to 10\%) and tracking (about 5\%) for the most central collisions. Other contributions are well below 1\%. For higher-order flow harmonics, the sine term contributes about 3\%  (13\%) for $\mathrm{v}_3$ ($\mathrm{v}_4$) for $|\eta|<2.5$. The MC closure at high $\eta$ ($|\eta|>1$) contributes up to 7\% (10\%) for $\mathrm{v}_3\{2\}$ ($\mathrm{v}_4\{2\}$) and less than 2\% for $\mathrm{v}_3\{4\}$ and $\mathrm{v}_4\{4\}$. For $|\eta|<1$ it is about 1\%. Other sources give contributions up to 4\%. 
\begin{table*}[th!] 
  \begin{center} 
      \caption[centEt]{Relative systematic and statistical uncertainties ($|\Delta 
\mathrm{v}_{2}|/\mathrm{v}_{2}$, in percent) for $\mathrm{v}_2$ measured with four-particle 
cumulants for three centrality intervals: 2--5\%, 15--20\% and 60--80\%.  A single entry is given where the uncertainty  only varies by a small amount over the selected $p_{\mathrm{T}}$ or $\eta$ range. Otherwise the range in uncertainties is provided corresponding to the range in $p_{\mathrm{T}}$ or $\eta$.} 
        \begin{tabular}{ll|rr|rr|rr} 
 \hline 
\multicolumn{2}{l|}{Centrality bin} & \multicolumn{2}{|c |}{2--5\% }& \multicolumn{2}{|c |}{15--20\%} &\multicolumn{2}{|c }{ 60--80\% }\\ \hline
 \multicolumn{2}{l|}{Measurement} & Syst.[\%] & Stat.[\%]&Syst.[\%] & Stat.[\%]&Syst.[\%] & Stat.[\%] \\ \hline
$\mathrm{v}_{2}\{4\}$ vs. $p_{\mathrm{T}}$ & $0.5 \leq p_{\mathrm{T}} \leq 1.5\,$\gev & 3.5 &2.9 & 5.8--1.2 &0.1 & 1.3--1.7 &1.2--1.4 \\
 & $1.5 < p_{\mathrm{T}} < 20\,$\gev & 3.5 &2.5--30 & 1.2--1.0 &0.1--41 &  1.8--5.4& 1.3--76 \\ \hline
$\mathrm{v}_{2}\{4\}$ vs. $\eta$ &    $|\eta| \leq 1$ & 3.8 &1.8 & 1.6 &0.1 & 2.4 &0.9 \\
 & $1 <|\eta| < 2.5$ & 6.0 &2.7 & 5.0 &0.1 & 2.3 &1.0 \\ \hline 
 $\mathrm{v}_{2}\{4\}$ vs. $N_{\mathrm{part}}$&  & 4.1 &0.5 & 3.0 &0.1 & 2.0 &0.3 \\ \hline 
        \end{tabular} 
    
        \label{tab:centETv2} 
  \end{center} 
\end{table*} 	 

The systematic uncertainty on the centrality dependence of $\mathrm{v}_2$ due to the centrality bin definition is 1--2\%. For the most central collisions,  the Monte Carlo closure gives a contribution of 4\%. The $r_0$ stability adds about 2\% only for $\mathrm{v}_2\{8\}$. Contributions from all other sources are below 1\%. For higher-order flow harmonics, the sine term and MC closure each  contribute about 3\%, while all other sources contribute less than 1\%.

The total systematic uncertainty is the sum in quadrature of all the individual contributions.  For illustration, Table~\ref{tab:centETv2} shows the total systematic uncertainties on $\mathrm{v}_2\{4\}$ for three representative centrality intervals: 2--5\%, 15--20\% and 60--80\%. For higher-order flow harmonics, the systematic uncertainties are listed in Table~\ref{tab:centETv3}.   For comparison, the statistical uncertainties on $\mathrm{v}_n$ are also listed.  It can be seen that the uncertainties on the measured  $\mathrm{v}_2$ at high \pT\, and on $\mathrm{v}_3$ and $\mathrm{v}_4$ over the whole kinematic range, are dominated by large statistical errors.
 \begin{table}[ht!] 
  \begin{center} 
          \caption[centEt]{Relative systematic and statistical uncertainties ($|\Delta 
\mathrm{v}_{n}|/\mathrm{v}_{n}$, in percent) for $\mathrm{v}_3$ and $\mathrm{v}_4$ measured with four-particle cumulants averaged over the accessible centrality ranges 0--60\% and 0--25\%, respectively. A single entry is given where the uncertainty  only varies by a small amount  over the  $p_{\mathrm{T}}$ range from 0.5 to 20~\gev\ or $\eta$ range from $-2.5$ to 2.5. Otherwise the range in uncertainties is provided corresponding to the range in $p_{\mathrm{T}}$ or $\eta$.} 
        \begin{tabular}{l|rr} 
 \hline 
 Measurement & Syst.[\%] & Stat.[\%] \\ \hline
$\mathrm{v}_{3}\{4\}$ vs. $p_{\mathrm{T}}$ & 6.2--4.8 &19--26 \\
$\mathrm{v}_{3}\{4\}$ vs. $\eta$ & 3.7 & 8  \\
$\mathrm{v}_{3}\{4\}$ vs. $N_{\mathrm{part}}$ &  3.3 & 16 \\ \hline
 $\mathrm{v}_{4}\{4\}$ vs. $p_{\mathrm{T}}$ & 16 & 46--34\\ 
$\mathrm{v}_{4}\{4\}$ vs. $\eta$ & 13 & 23\\ 
$\mathrm{v}_{4}\{4\}$ vs. $N_{\mathrm{part}}$ &  5.4 & 31 \\ \hline 
        \end{tabular} 
        \label{tab:centETv3} 
  \end{center} 
\end{table} 	 

In addition to the evaluation of the systematic uncertainty, further cross-checks are performed. The comparison between cumulants calculated with the GFC and QC methods is discussed in detail in Sect.~\ref{sec:analysis}. The analysis is performed separately for negatively and positively charged particles and the resulting $\mathrm{v}_n$ coefficients are found to be consistent within their statistical and systematic uncertainties. Furthermore, the consistency of $\mathrm{v}_2\{2k\}$ for $k>1$ measured for same-sign particles and all combinations of charged particles confirms the global collective feature of the measured effect. Consistency is also observed between measurements obtained from sub-samples collected early and late during the data-taking period. The analysis also evaluates a potential bias which may be due to the large spread in charged-particle multiplicities in centrality intervals defined  by  {\mbox{$\Sigma E_{\mathrm{T}}^{\mathrm{FCal}}$}}. For this purpose, in a given centrality bin selected by {\mbox{$\Sigma E_{\mathrm{T}}^{\mathrm{FCal}}$}}, the analysis is restricted to events with multiplicities limited to a very narrow range  (corresponding to $\pm$RMS/2 around the mean multiplicity) and compared to the analysis performed for the full range of multiplicities.  Both give $\mathrm{v}_n$ harmonics consistent with each other within their statistical and systematic uncertainties. 
\section{Results}
\label{sec:results}
\subsection{Transverse momentum dependence of flow harmonics}
\label{sec:pTdep}
\begin{figure*}[ht!]
\begin{center}
 \includegraphics[width=150mm]{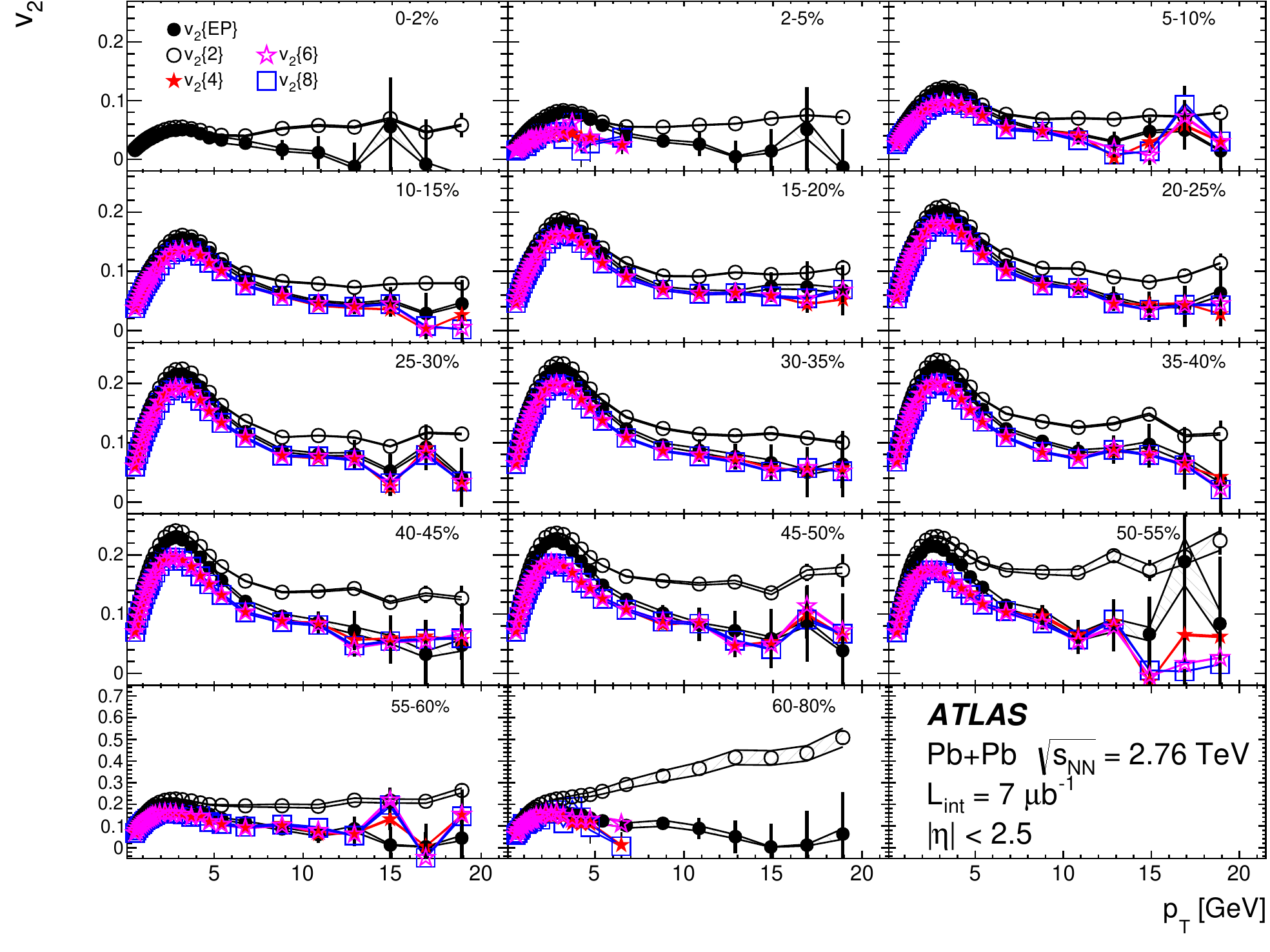}
    \caption{The second flow harmonic calculated with the two-, four-, six-, and eight-particle cumulants measured over the full pseudorapidity range, $|\eta|<2.5$, as a function of transverse momentum in different centrality intervals, indicated on the plots.  For the most central collisions (0--2\% centrality class) the results are available only for $\mathrm{v}_2\{2\}$. For comparison the $\mathrm{v}_2\{\mathrm{EP}\}$ measurements obtained with the event-plane method are also shown. Statistical uncertainties are shown as bars  and systematic uncertainties as bands.
}
\label{fig:Fig5}
\end{center}
\end{figure*}
\begin{figure*}[ht!]
\begin{center}
\includegraphics[width=149mm]{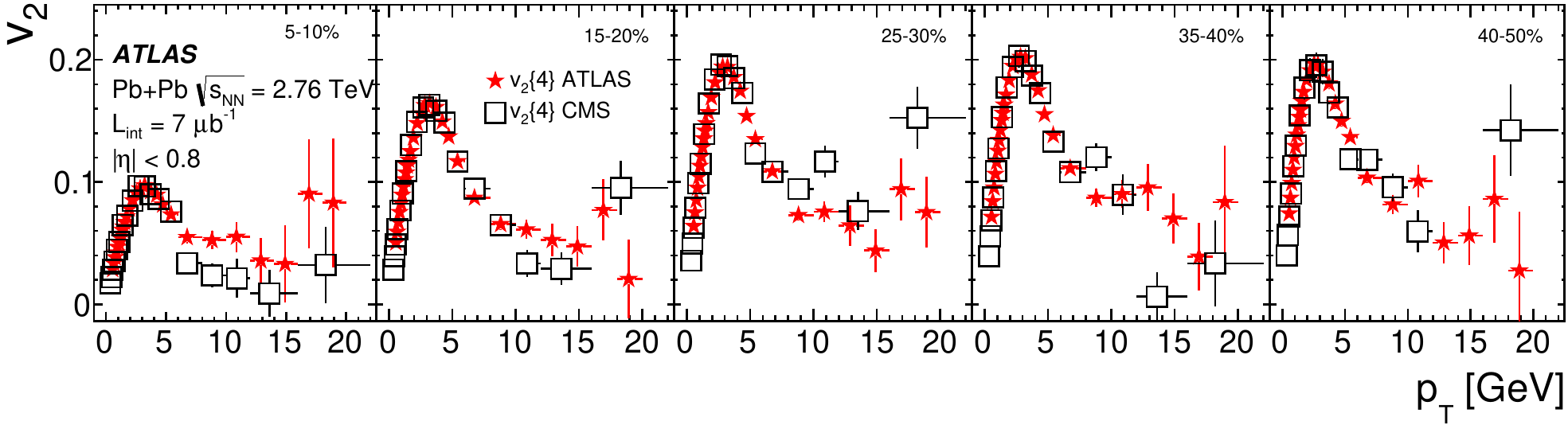}
\includegraphics[width=149mm]{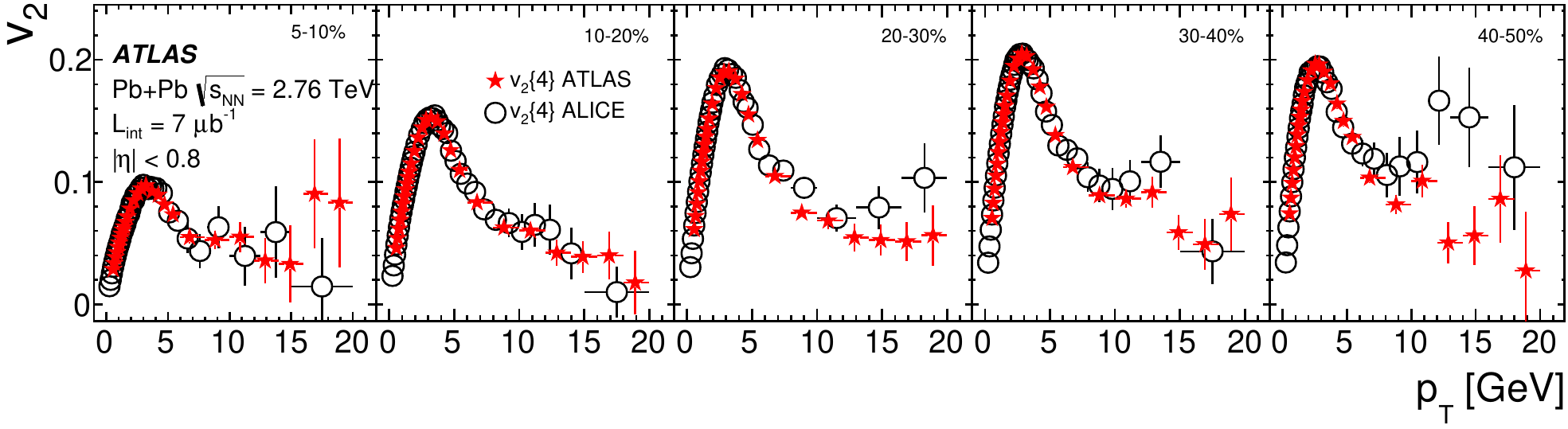}
    \caption{Comparison of the ATLAS and CMS \citep{Cms1} (top panel), and ATLAS and ALICE \citep{Alice3} (bottom panel) measurements of $\mathrm{v}_2\{4\}$ for selected centrality intervals at $|\eta|<0.8$. The error bars denote statistical and systematic uncertainties added in quadrature.  
}
\label{fig:Fig6}
\end{center}
\end{figure*}
\begin{figure*}[ht!]
\begin{center}
\includegraphics[height=57mm]{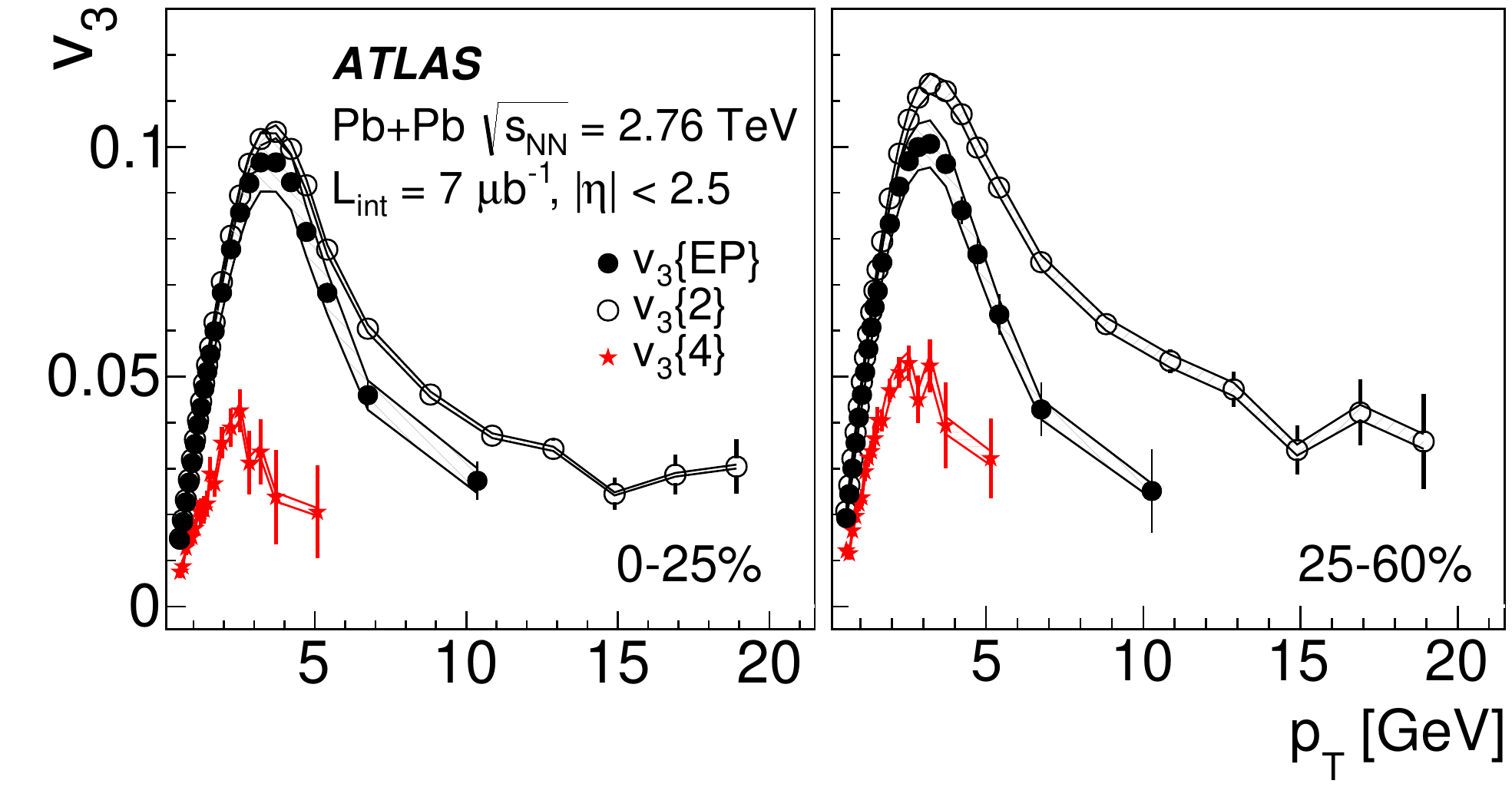}
\includegraphics[height=57mm]{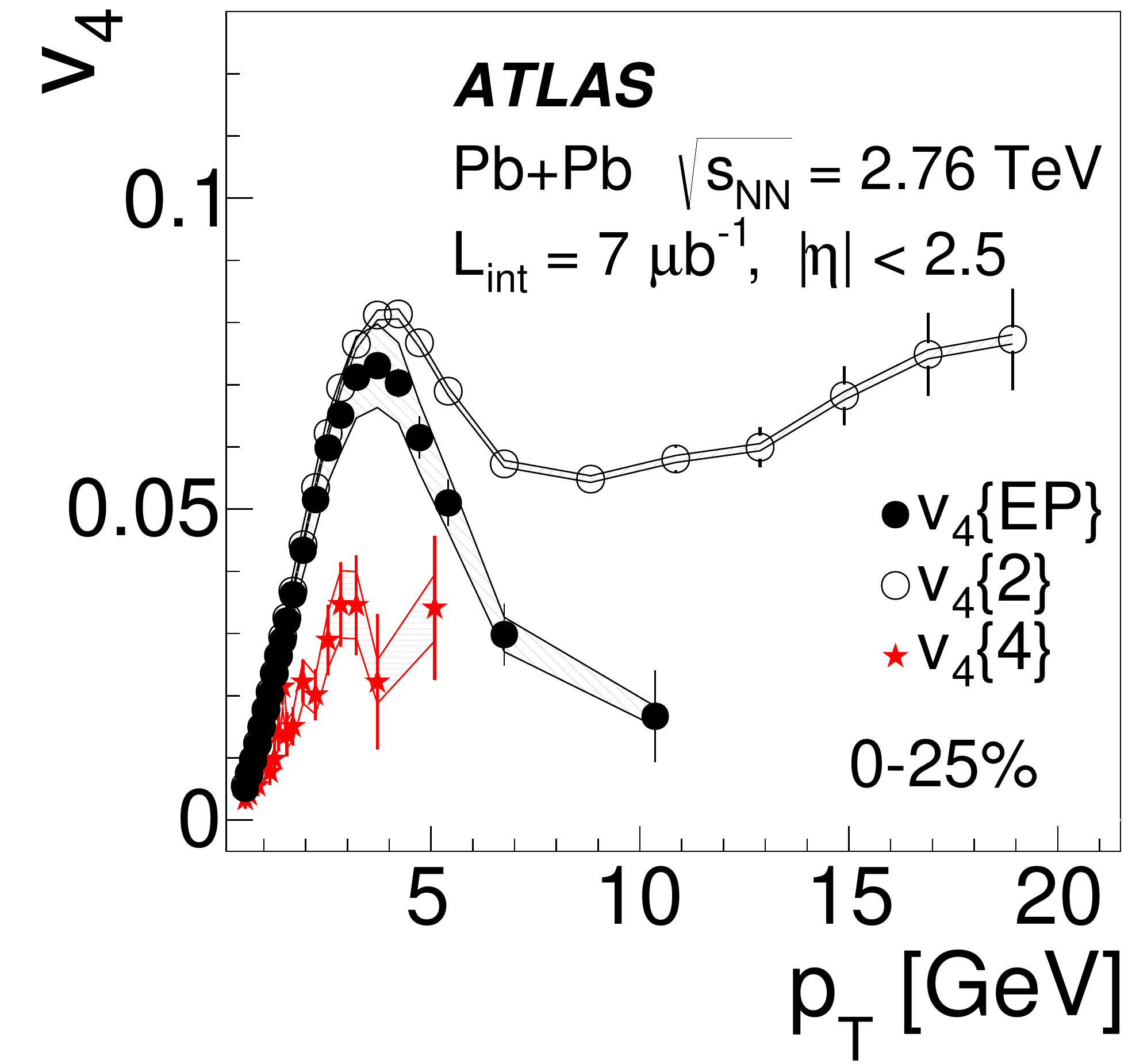}
    \caption{The transverse momentum dependence of $\mathrm{v}_3$ calculated with two- and four-particle cumulants and with the event-plane method $\mathrm{v}_3\{\mathrm{EP}\}$ for the centrality interval 0--25\% (left plot) and 25--60\% (middle plot). The right plot shows the same results for $\mathrm{v}_4$ for the centrality interval 0--25\%. Statistical errors are shown as bars and systematic uncertainties as bands. The highest \pT\ measurement for $\mathrm{v}_n\{4\}$ ($\mathrm{v}_n\{\mathrm{EP}\}$) is integrated over the \pT\   range 4--20 (8--20)~\gev.}
\label{fig:Fig7}
\end{center}
\end{figure*}
The transverse momentum dependence of $\mathrm{v}_2\{2\}$, $\mathrm{v}_2\{4\}$, $\mathrm{v}_2\{6\}$ and $\mathrm{v}_2\{8\}$ is shown  in Fig.~\ref{fig:Fig5} in 14 centrality intervals as indicated in the plots. The $\mathrm{v}_2$ coefficients are integrated over the full pseudorapidity range $|\eta|<2.5$.  The elliptic flow measurements, $\mathrm{v}_2\{\mathrm{EP}\}$, obtained with the event-plane method \citep{Poskanzer:1998yz} are also shown. For this comparison, the measurements from Ref.~\citep{Atlas1} were reanalysed with the same track-quality requirements and centrality intervals as for the present analysis. 
The event-plane $\mathrm{v}_2$ is systematically smaller than $\mathrm{v}_2\{2\}$ since it is less affected by short-range two-particle correlations, which are partially removed from $\mathrm{v}_2\{\mathrm{EP}\}$ due to the separation between the phase-space region where the event plane angle is determined ($3.2<|\eta|<4.9$) and the phase space where charged-particle momenta are reconstructed ($|\eta|<2.5$). This effect is particularly pronounced at high transverse momenta, where $\mathrm{v}_2\{2\}$ is strongly influenced by jet-related correlations. At lower transverse momenta, the difference between $\mathrm{v}_2\{2\}$ and $\mathrm{v}_2\{\mathrm{EP}\}$ can also be attributed to flow fluctuations. The difference between $\mathrm{v}_2\{\mathrm{EP}\}$ and $\mathrm{v}_2\{4\}$ is mainly due to flow fluctuations.
The $\mathrm{v}_2\{2k\}$ for $k>1$ is systematically smaller than $\mathrm{v}_2\{2\}$, consistent with the expected suppression of non-flow effects in $\mathrm{v}_2$ obtained with cumulants of more than two particles.  The results for $\mathrm{v}_2\{2k\}$ with $k>1$ agree with each other, within the uncertainties, for all centrality intervals, indicating that already  the four-particle cumulants efficiently suppress non-flow correlations.  As a function of transverse momentum, the second flow harmonic first increases with $p_{\mathrm{T}}$ up to $p_{\mathrm{T}}\approx$ 2--3~\gev, then gradually decreases for \pT\ values up to about 6~\gev. This trend is consistent with hydrodynamic predictions for a collective expansion of the created system \citep{hydropT1,hydropT2}. Beyond $\pT$ of about 10~\gev, a much weaker $\mathrm{v}_2$ dependence on \pT\ is observed. Interestingly the coefficients $\mathrm{v}_2\{2k\}$ for $k>1$ remain significant  at high transverse momenta, up to about 20~\gev, over a broad centrality range, except the most peripheral and the most central collisions. These large values of $\mathrm{v}_2\{4\}$, $\mathrm{v}_2\{6\}$ and $\mathrm{v}_2\{8\}$ at high transverse momenta may reflect
both the anisotropy of the initial geometry and the path-length dependence of the parton energy loss in the dense, strongly interacting medium \citep{atlas_jetv2}.

Figure~\ref{fig:Fig6} shows the comparison of our results for $\mathrm{v}_2\{4\}$ integrated over $|\eta|<0.8$ as a function of \pT,  to  these coefficients measured by the CMS \citep{Cms1} and ALICE \citep{Alice3} experiments in several centrality intervals. The results on the elliptic flow harmonic measured with four-particle cumulants are consistent within uncertainties for the three experiments. 

The transverse momentum dependence of the higher-order harmonics, $\mathrm{v}_3$ and $\mathrm{v}_4$, is shown in Fig.~\ref{fig:Fig7}  and compared to the results obtained with the event-plane method. Due to the large uncertainties on the harmonics measured with four-particle cumulants, especially for events with low multiplicities, the results are shown in wide centrality ranges: for $\mathrm{v}_3$ in the two broad centrality intervals, 0--25\% and 25--60\%, and for $\mathrm{v}_4$ in the full accessible centrality range, 0--25\%.  In addition, the results for $\mathrm{v}_n\{4\}$ are shown in fine \pT\ bins at low transverse momenta, up to 4~\gev, while the last \pT\ point covers the range from 4~\gev\ to 20~\gev. Similarly to $\mathrm{v}_2$, smaller short-range jet-like correlations are observed in $\mathrm{v}_{3,4}\{\mathrm{EP}\}$ as compared to $\mathrm{v}_{3,4}\{2\}$. Significantly non-zero values of the third and fourth flow harmonics calculated with four-particle cumulants are observed with a \pT\ dependence similar to that of $\mathrm{v}_2$. The $\mathrm{v}_n\{4\}$ harmonic is systematically smaller than $\mathrm{v}_n\{2\}$, consistent with the suppressed non-flow effects in flow harmonics obtained with cumulants of more than two particles. It is noted that the difference between $\mathrm{v}_3\{4\}$ ($\mathrm{v}_4\{4\}$) and $\mathrm{v}_3\{\mathrm{EP}\}$ ($\mathrm{v}_4\{\mathrm{EP}\}$), which amounts to a factor of about two, is much larger than the difference between  $\mathrm{v}_2\{4\}$ and $\mathrm{v}_2\{\mathrm{EP}\}$, which is of the order of 30\%. This indicates that fluctuations of higher-order flow harmonics are much stronger than fluctuations of $\mathrm{v}_2$.
\subsection{Pseudorapidity dependence of flow harmonics}
\label{sec:etadep}
\begin{figure*}[ht!]
\begin{center}
\includegraphics[width=145mm]{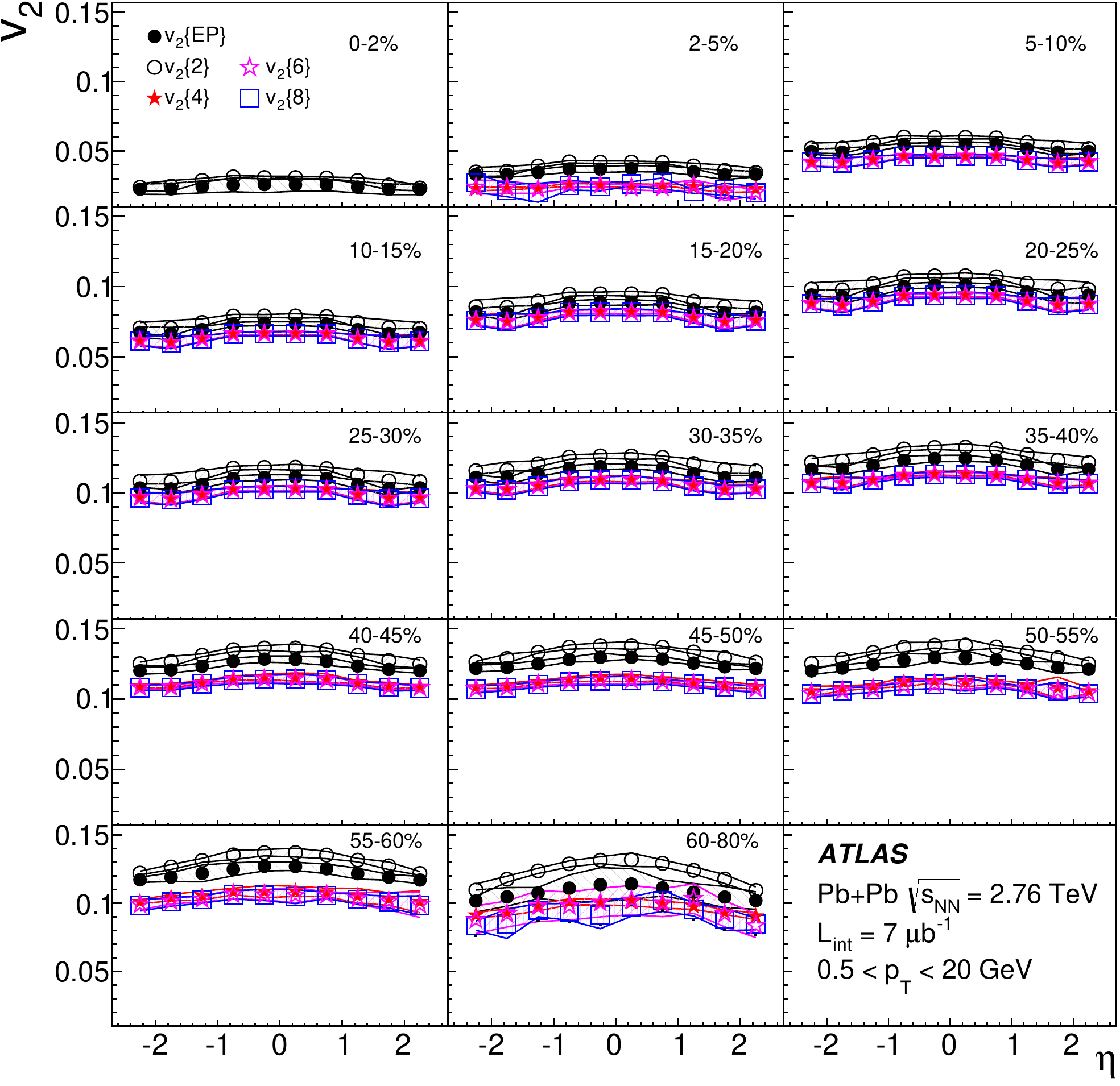}
    \caption{The second flow harmonic calculated with the two-, four-, six-, and eight-particle cumulants as a function of $\eta$  in different centrality intervals, integrated over the \pT\ range $0.5 < p_{\mathrm{T}}<20 \,$\gev. The results for $\mathrm{v}_2\{\mathrm{EP}\}$ are also shown. Statistical errors are shown as bars (too small to be seen on this scale) and systematic uncertainties as shaded bands. 
}
\label{fig:Fig9}
\end{center}
\end{figure*}
\begin{figure*}[ht!]
\begin{center}
\includegraphics[width=70mm]{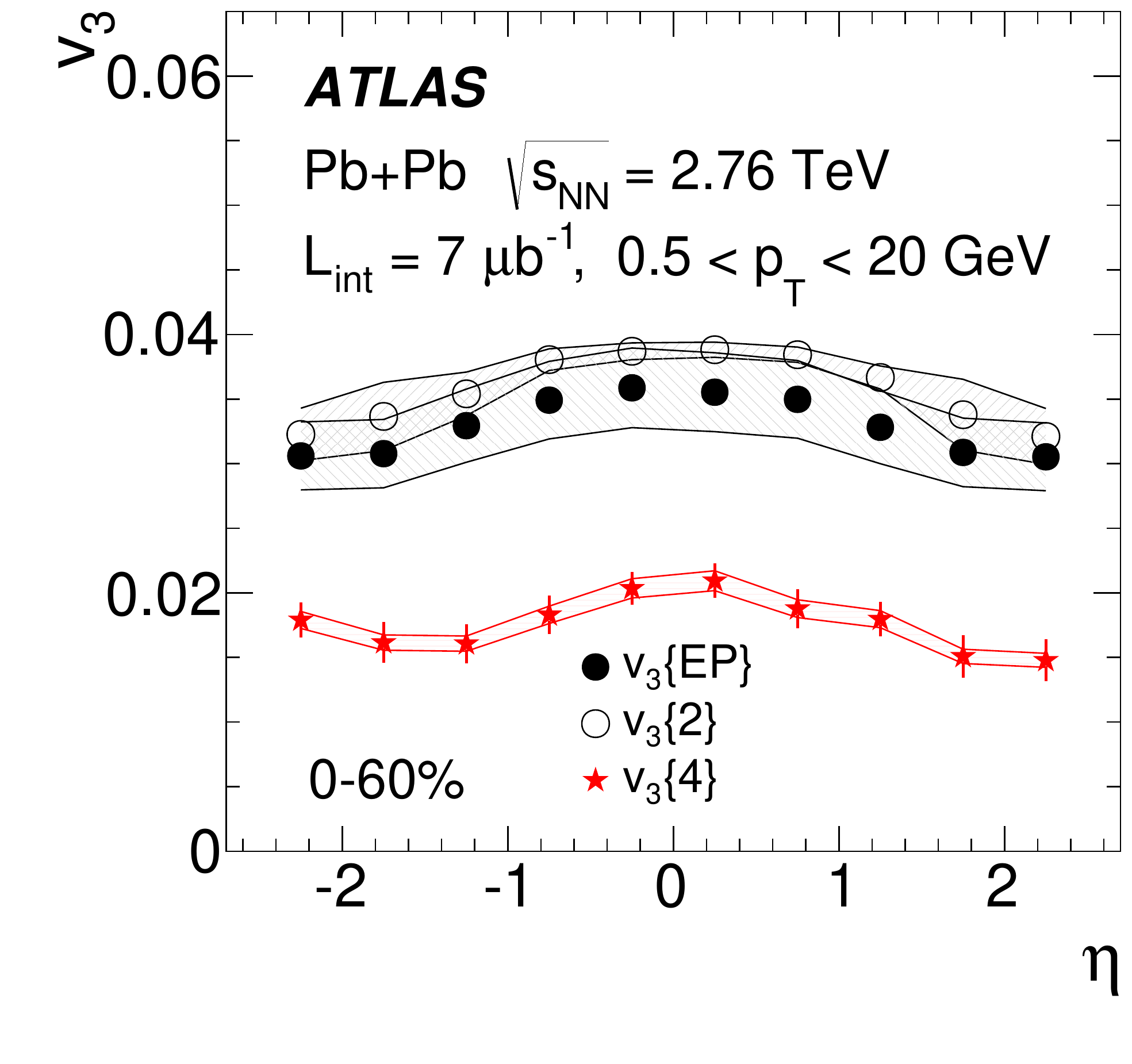}
\includegraphics[width=70mm]{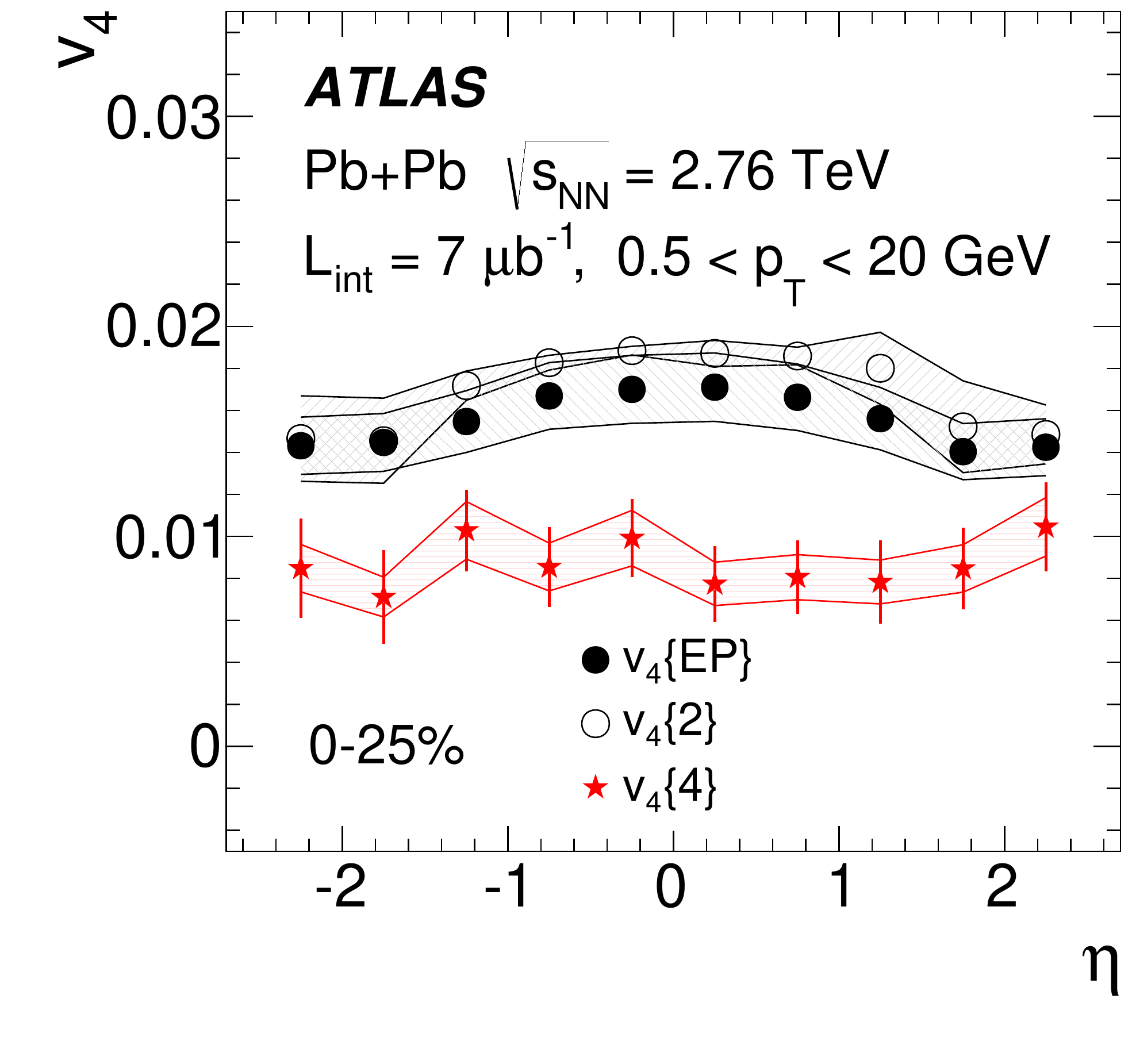}
    \caption{The pseudorapidity dependence of $\mathrm{v}_3$ (left plot) calculated with two- and four-particle cumulants and with the event-plane method, integrated over \pT\ from 0.5 \gev\  to 20 \gev\ for the centrality  range 0--60\%. The same is shown for $\mathrm{v}_4$ (right plot) for the centrality range 0--25\%.  Statistical errors are shown as bars and systematic uncertainties as shaded bands. 
}
\label{fig:Fig10}
\end{center}
\end{figure*}
The pseudorapidity dependence of $\mathrm{v}_n\{2k\}$ is studied as a function of centrality for flow coefficients integrated over the \pT\ range from 0.5~\gev\ to 20~\gev. Figure~\ref{fig:Fig9} shows $\mathrm{v}_2\{2\}$, $\mathrm{v}_2\{4\}$, $\mathrm{v}_2\{6\}$, $\mathrm{v}_2\{8\}$ and $\mathrm{v}_2\{\mathrm{EP}\}$ as a function of $\eta$ in 14 centrality intervals as indicated in the plots. Observations similar to the case of the \pT\ dependence can be made: $\mathrm{v}_2\{2k\}$ for $k>1$ is systematically smaller than $\mathrm{v}_2\{2\}$ and $\mathrm{v}_2\{\mathrm{EP}\}$, while the results for $\mathrm{v}_2\{2k\}$ with $k>1$ agree with each other for all centrality intervals. No strong dependence on pseudorapidity is observed for any of the second flow harmonic measurements in any of the centrality bins. Some weak dependence is observed only for $\mathrm{v}_2\{2\}$ and can be attributed to the contributions from short-range two-particle correlations. A weak pseudorapidity dependence is observed for $\mathrm{v}_3\{4\}$ as shown in Fig.~\ref{fig:Fig10} for harmonics averaged over the full accessible centrality range (0--60\%). The fourth-order flow harmonics, $\mathrm{v}_4\{4\}$, show no significant dependence on pseudorapidity, within the measurement uncertainties, over the centrality range 0--25\%. A systematic reduction in the non-flow contribution is observed for $\mathrm{v}_n\{\mathrm{EP}\}$ as compared to  $\mathrm{v}_n\{2\}$.

\subsection{Centrality dependence of the integrated flow harmonics}
\label{sec:centdep}
\begin{figure*}[ht!]
\begin{center}
\includegraphics[width=70mm]{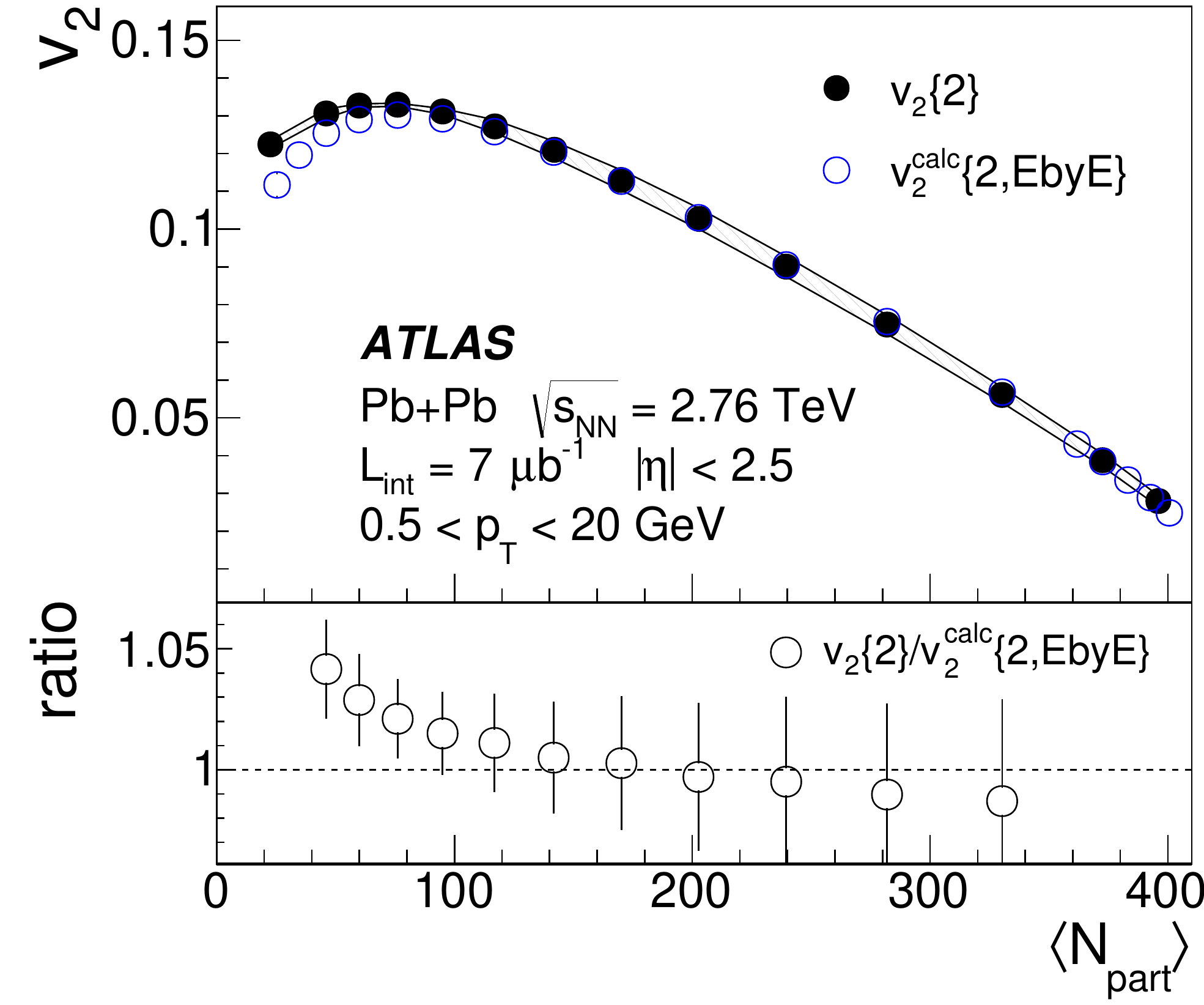}
\includegraphics[width=70mm]{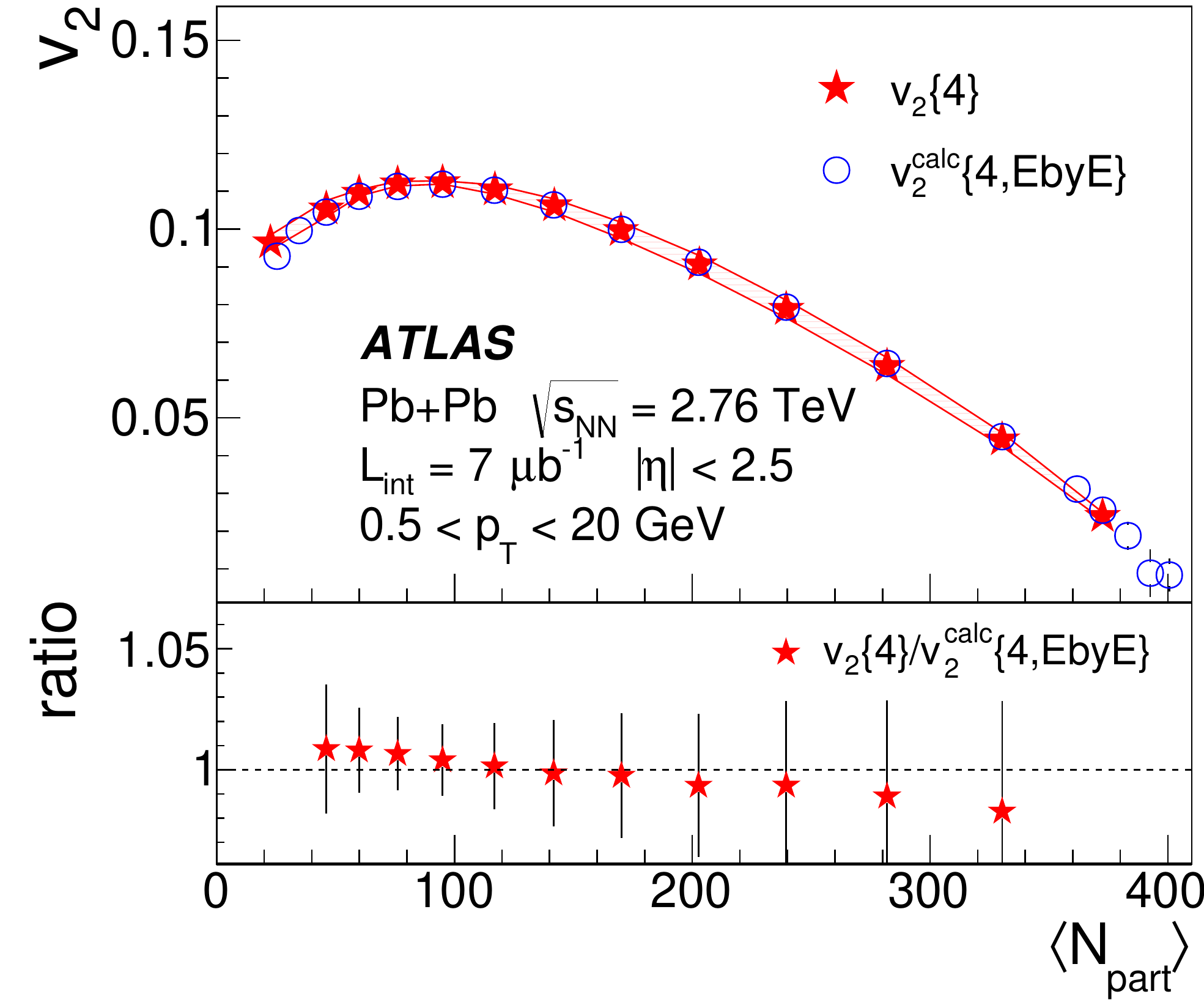}
\includegraphics[width=70mm]{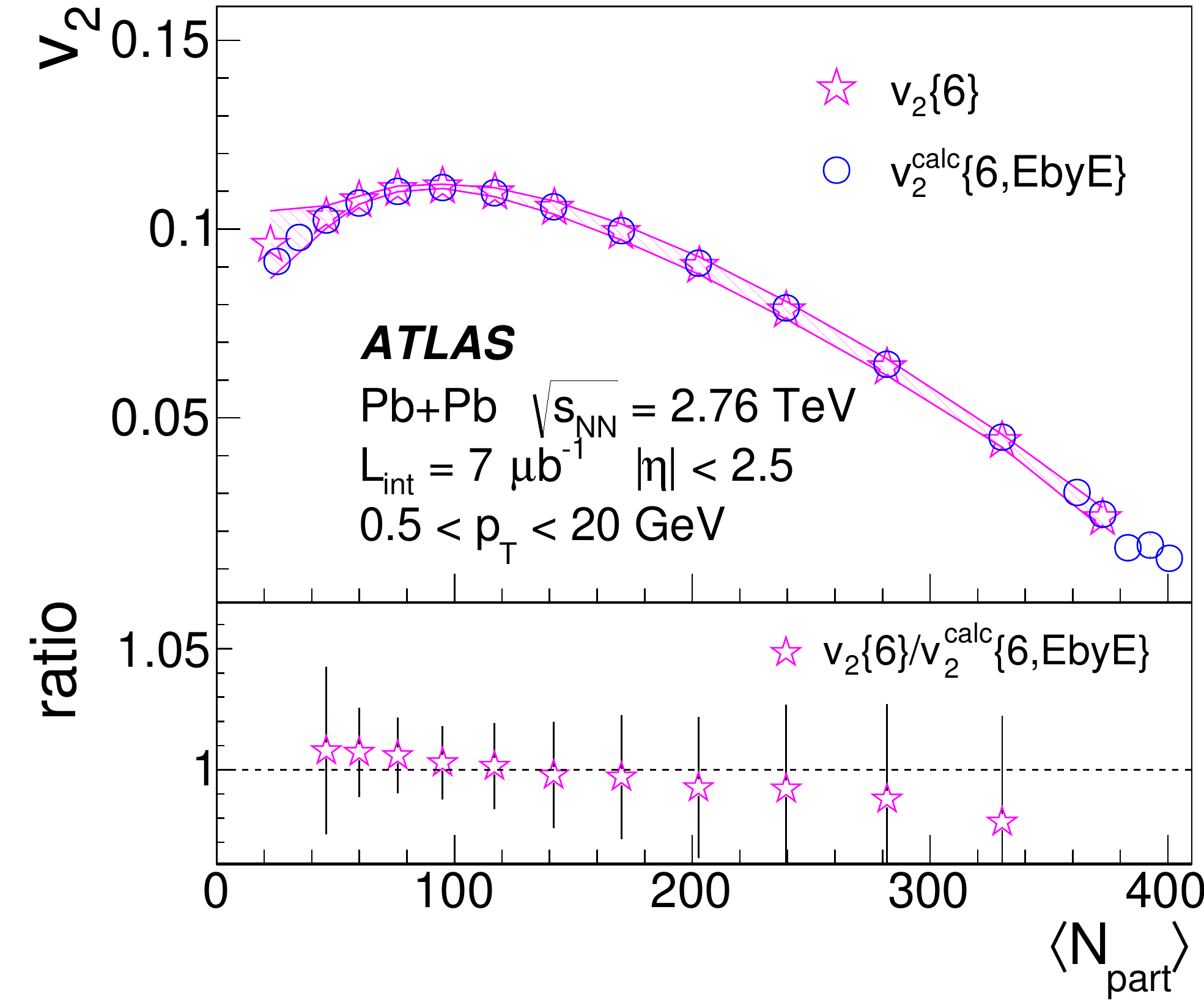}
\includegraphics[width=70mm]{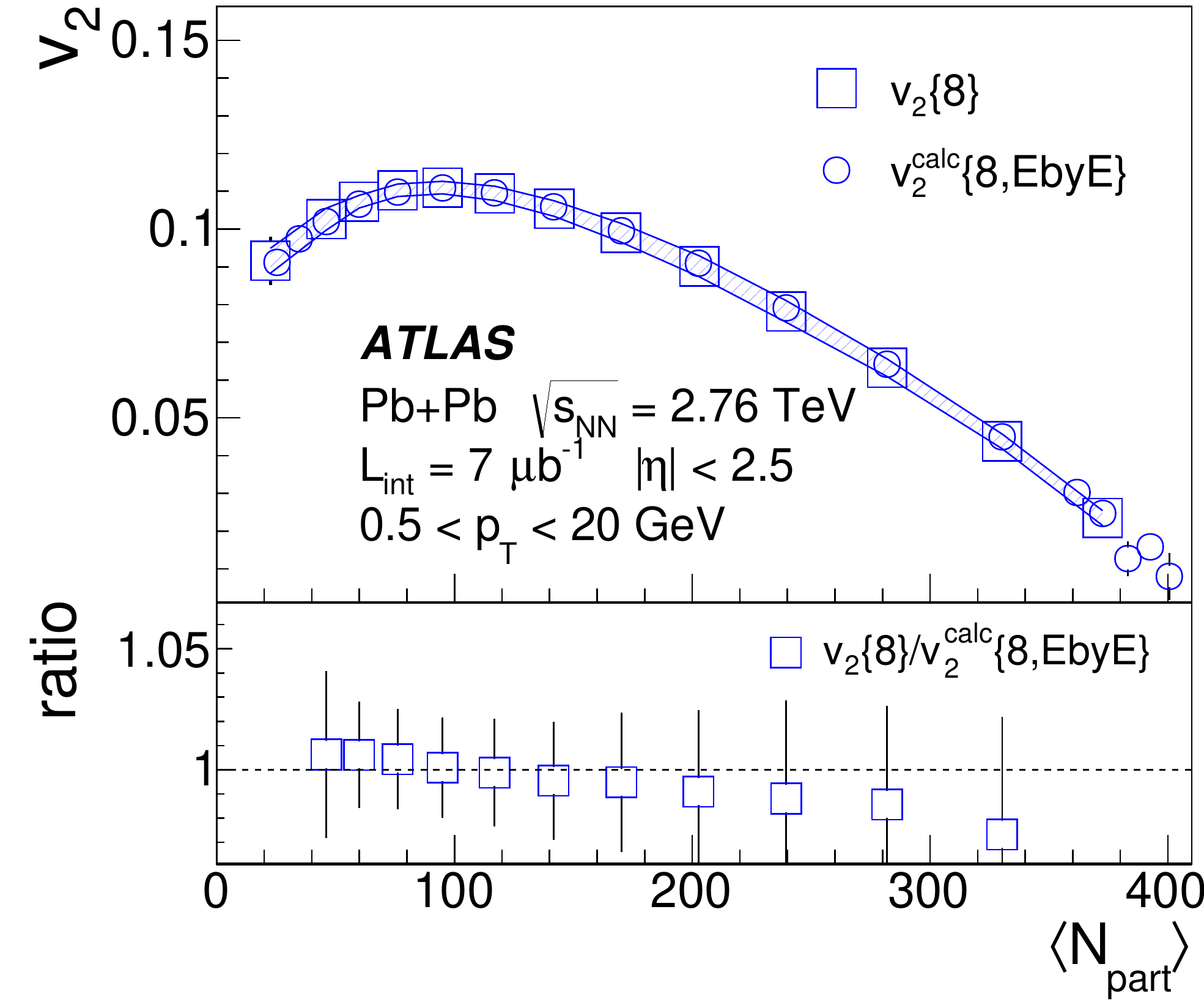}
    \caption{Comparison of the centrality dependence of $\mathrm{v}_2\{2k\}$ for $k=$1--4 integrated over \pT\ from 0.5 \gev\ to 20 \gev\ and over $|\eta|<2.5$, and $\mathrm{v}_2^{\mathrm{calc}}\{2k,\mathrm{EbyE}\}$ calculated from the measured $\mathrm{v}_2$ distribution \citep{Atlas3}.  For $\mathrm{v}_2\{2k\}$, the statistical errors are shown as bars and the systematic uncertainties as shaded bands. For $\mathrm{v}_2^{\mathrm{calc}}\{2k,\mathrm{EbyE}\}$, the error bars denote statistical and systematic errors added in quadrature.  The bottom panels in each plot show ratios  of the results obtained with the two methods. The ratios are calculated for matching centrality intervals.
}
\label{fig:Fig11}
\end{center}
\end{figure*}
\begin{figure*}[ht!]
\begin{center}
\includegraphics[width=70mm]{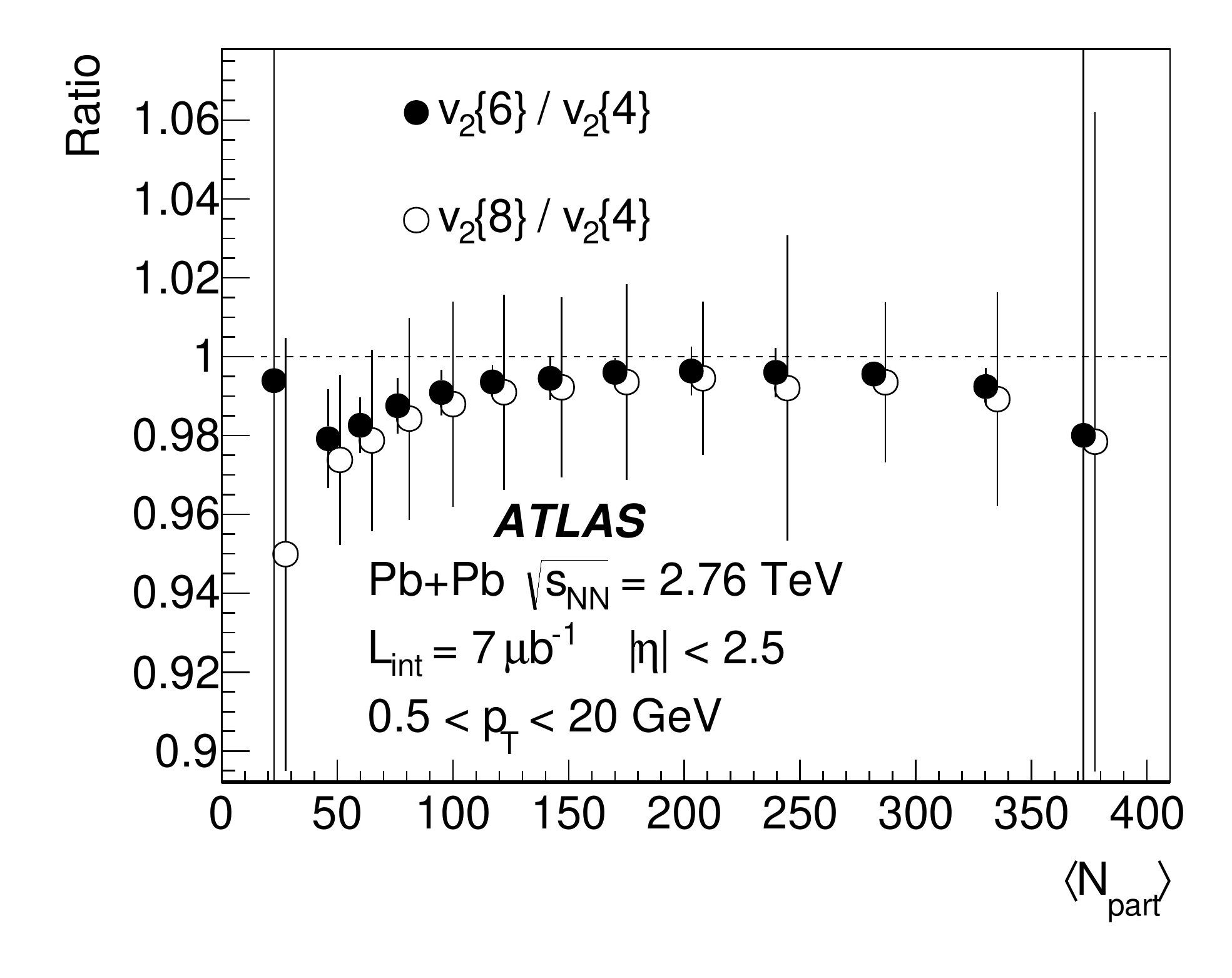}
\includegraphics[width=70mm]{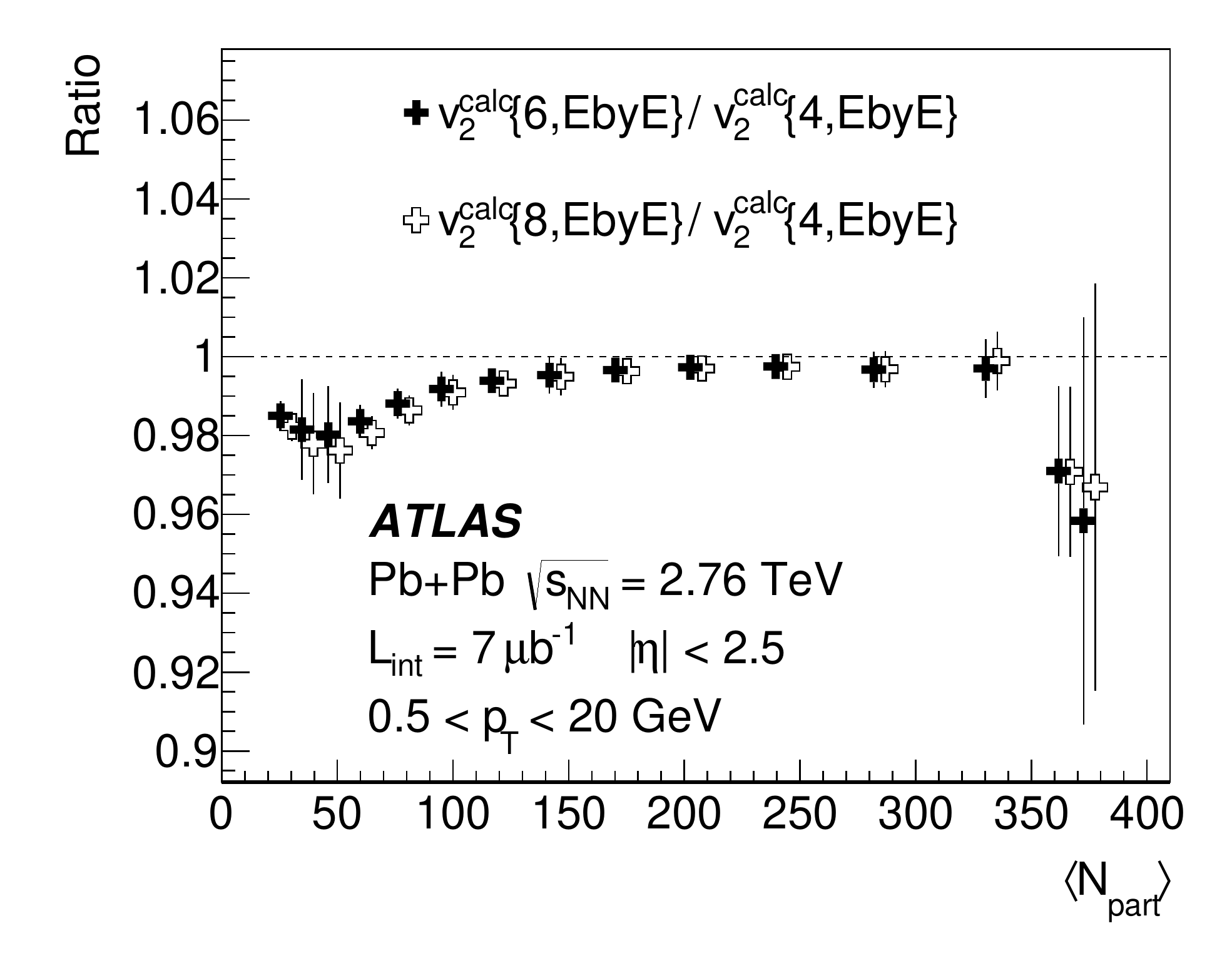}
    \caption{The ratio of $\mathrm{v}_2\{6\}$ and $\mathrm{v}_2\{8\}$ to $\mathrm{v}_2\{4\}$ as a function of the average number of participating nucleons, $\langle N_{\mathrm{part}}\rangle$, for elliptic flow coefficients obtained from the cumulant method (left) and calculated from the measured $p(\mathrm{v}_2)$ distribution \citep{Atlas3} (right).  The error bars denote statistical and systematic errors added in quadrature.  The ratio symbols are shifted horizontally with respect to each other for clarity.}
\label{fig:Fig12}
\end{center}
\end{figure*}
\begin{figure*}[ht!]
\begin{center}
\includegraphics[width=70mm]{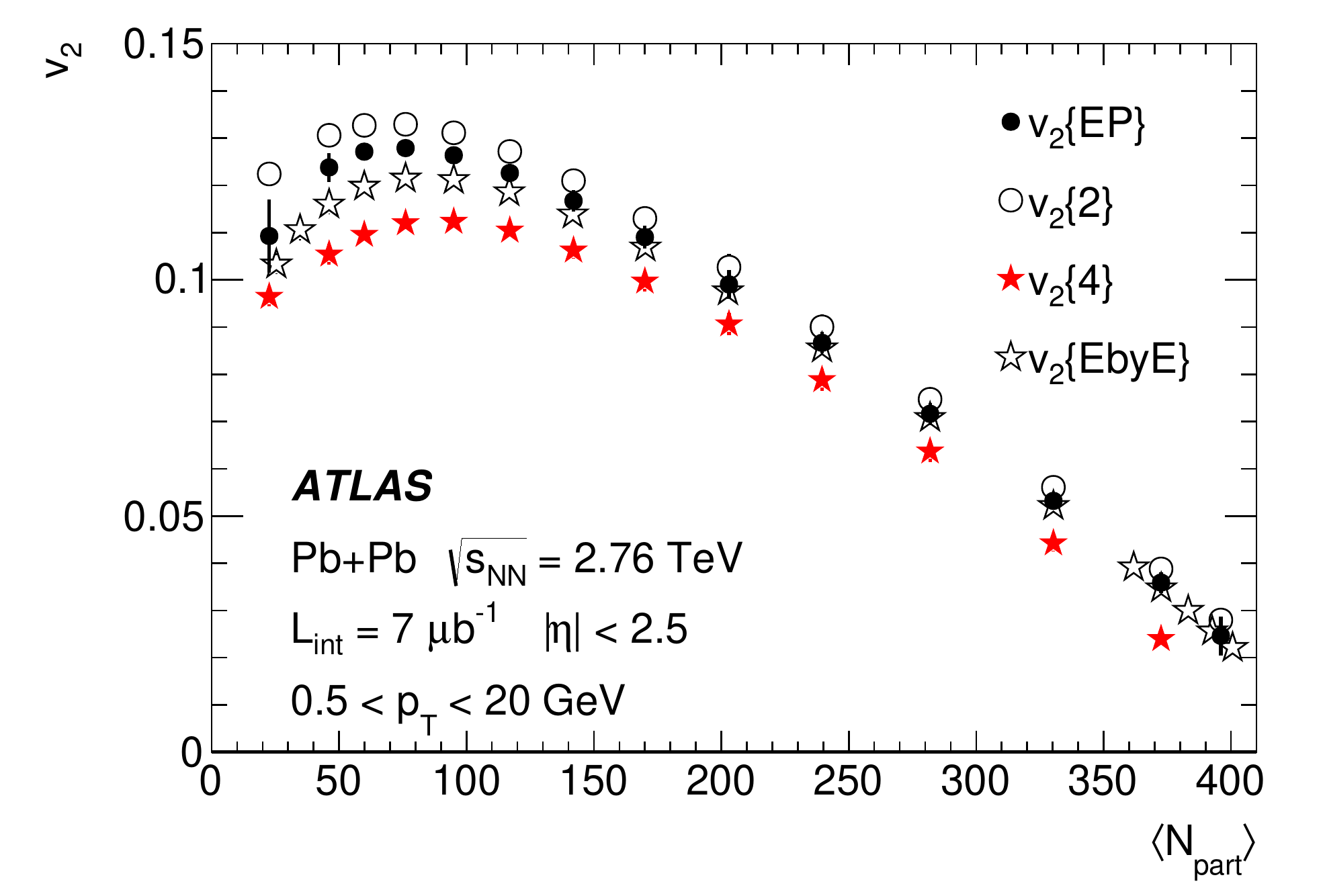}
\includegraphics[width=70mm]{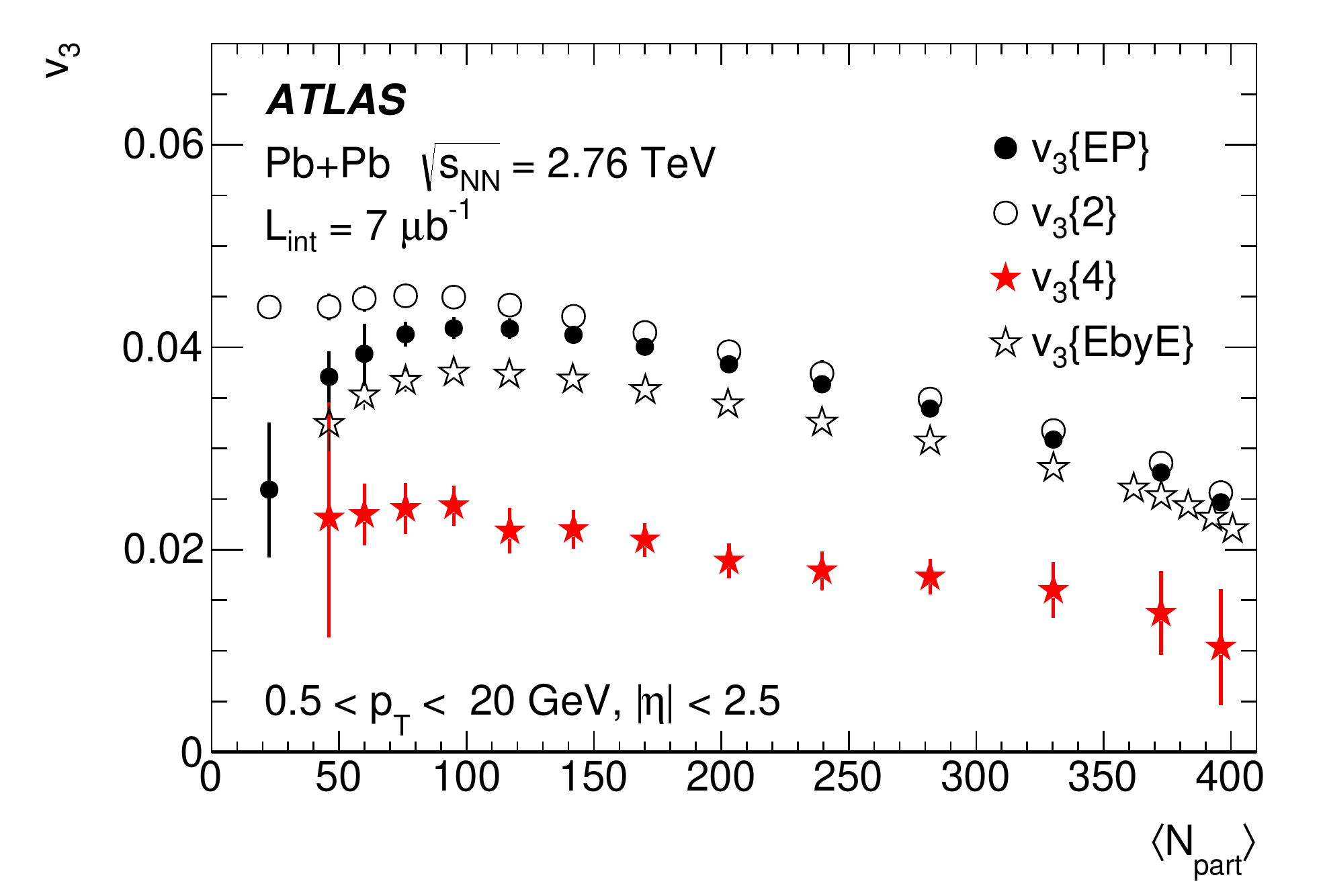}
\includegraphics[width=70mm]{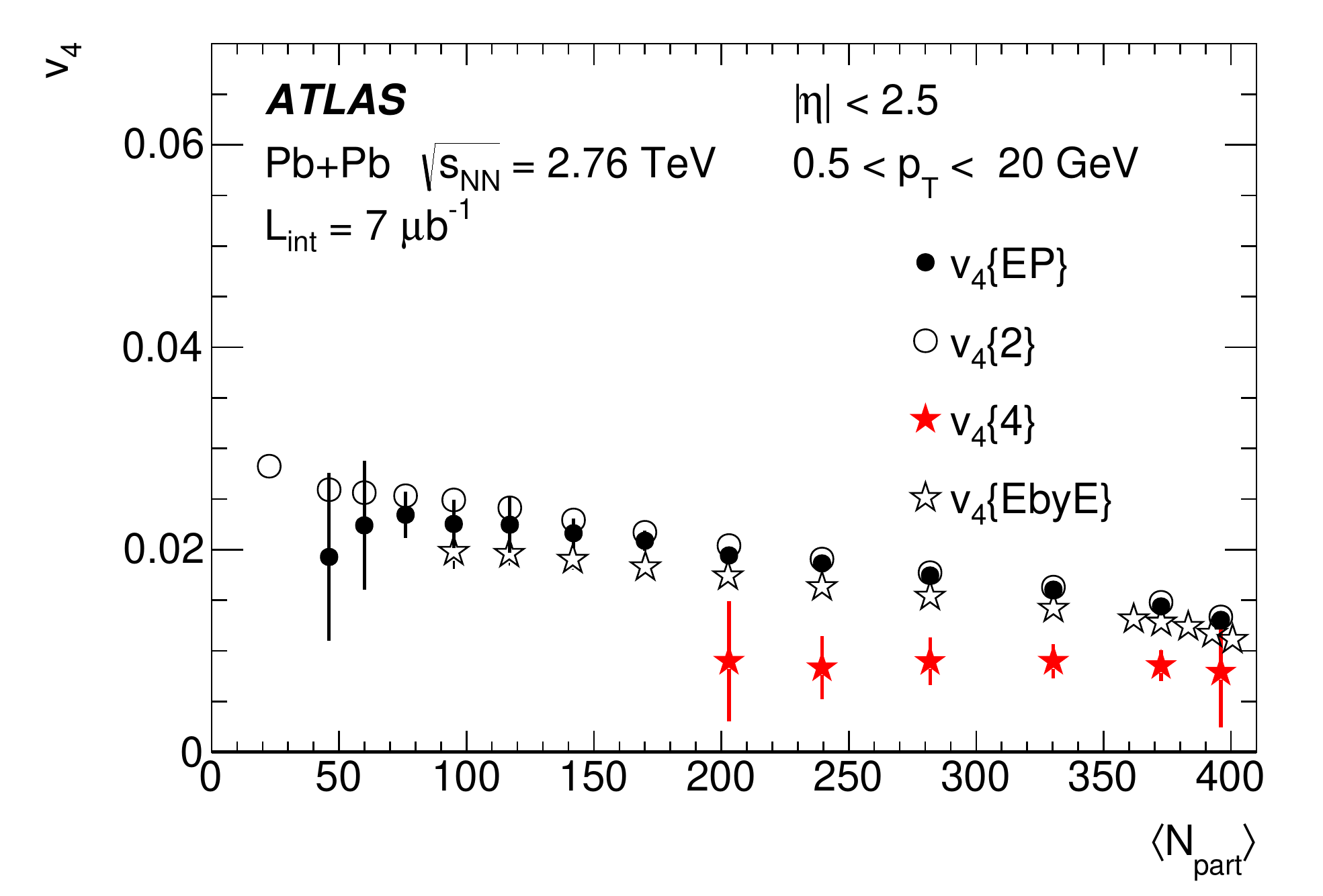}
    \caption{Comparison of the $\langle N_{\mathrm{part}}\rangle$ dependence of the $\mathrm{v}_2$ (top left), $\mathrm{v}_3$ (top right) and $\mathrm{v}_4$ (bottom) harmonics measured with different methods, with $\mathrm{v}_n\{\mathrm{EbyE}\}$ denoting the mean value of the corresponding $p(\mathrm{v}_n)$. The error bars denote statistical and systematic uncertainties added in quadrature. 
}
\label{fig:Fig12a}
\end{center}
\end{figure*}
The centrality dependence of the elliptic flow harmonic, integrated over the full range in $\eta$ and \pT\ and obtained with cumulants of various orders, is shown as a function of $\langle N_{\mathrm{part}}\rangle$ in Fig.~\ref{fig:Fig11}. The coefficients $\mathrm{v}_2\{2k\}$, and in general $\mathrm{v}_n\{2k\}$, can also be calculated from the moments of the distribution, $p(\mathrm{v}_n)$, of the event-by-event ($\mathrm{EbyE}$) measured flow harmonics, $\langle \mathrm{v}_n^k\rangle=\sum\mathrm{v}_n^k p(\mathrm{v}_n)$ as:
{\begin{eqnarray}
(\mathrm{v}_n^{\mathrm{calc}}\{2,\mathrm{EbyE}\})^2 &\equiv &\langle \mathrm{v}_n^2\rangle,   \\
(\mathrm{v}_n^{\mathrm{calc}}\{4,\mathrm{EbyE}\})^4 &\equiv & -\langle \mathrm{v}_n^4 \rangle +2\langle \mathrm{v}_n^2 \rangle ^2,  \\
(\mathrm{v}_n^{\mathrm{calc}}\{6,\mathrm{EbyE}\})^6 &\equiv & \left( \langle \mathrm{v}_n^6 \rangle -9\langle \mathrm{v}_n^4\rangle \langle \mathrm{v}_n^2\rangle +12 \langle \mathrm{v}_n^2\rangle ^3\right)/4, \\
(\mathrm{v}_n^{\mathrm{calc}}\{8,\mathrm{EbyE}\})^8 &\equiv &-\left( \langle \mathrm{v}_n^8\rangle -16\langle \mathrm{v}_n^6\rangle \langle \mathrm{v}_n^2 \rangle -18 \langle \mathrm{v}_n^4\rangle ^2 \right) /33\nonumber \\
 & &  -\left(144 \langle \mathrm{v}_n^4\rangle \langle \mathrm{v}_n^2\rangle ^2 -144\langle \mathrm{v}_n^2\rangle^4 \right) /33. 
\end{eqnarray}}
 \noindent
ATLAS has measured $p(\mathrm{v}_n)$ for $n=2,3,4$ \citep{Atlas3} . 
The comparison of $\mathrm{v}_2\{2k\}$ obtained with the cumulant method to  $\mathrm{v}_2^{\mathrm{calc}}\{2k,\mathrm{EbyE}\}$ is shown in Fig.~\ref{fig:Fig11}. Good agreement between the two independent measurements is seen. The cumulant method gives $\mathrm{v}_2$ values larger than those calculated from the $p(\mathrm{v}_2)$ distribution only for $\mathrm{v}_2\{2\}$ measured in the most peripheral collisions, due to contributions from short-range two-particle correlations in the former. The ratios of $\mathrm{v}_2\{6\}$ and $\mathrm{v}_2\{8\}$ to $\mathrm{v}_2\{4\}$ are shown in Fig.~\ref{fig:Fig12}. The left panel shows results from the cumulant method. The ratios are systematically below unity, most significantly at low $N_{\mathrm{part}}$.  This effect, which is of the order of 1--2\%, is significant for the ratio $\mathrm{v}_2\{6\}/\mathrm{v}_2\{4\}$ while it is within the present uncertainty of the cumulant measurements for $\mathrm{v}_2\{8\}/\mathrm{v}_2\{4\}$. Better precision is achieved for $\mathrm{v}_2^{\mathrm{calc}}\{2k,\mathrm{EbyE}\}$ (right panel of Fig.~\ref{fig:Fig12}).  The difference between $\mathrm{v}_2\{4\}$ and $\mathrm{v}_2\{6\}$ or $\mathrm{v}_2\{8\}$ is attributed to the non-Bessel-Gaussian character of the $p(\mathrm{v}_2)$ distribution measured in peripheral collisions \citep{Atlas3}. 

It is interesting to compare flow harmonic measurements obtained with different methods, which have different sensitivities to non-flow correlations and flow harmonic fluctuations. Since the higher-order flow harmonics, $\mathrm{v}_n\{2k\}$ ($n>2$), are measured with the cumulant method with up to four-particle cumulants (see Sect.~\ref{sec:analysis}), the $\mathrm{v}_n\{2\}$ and $\mathrm{v}_n\{4\}$ are only included in the comparison. 
Figure~\ref{fig:Fig12a} shows the comparison of $\mathrm{v}_2$, $\mathrm{v}_3$ and $\mathrm{v}_4$ obtained using the cumulant method with the ATLAS results obtained with the event-plane method, $\mathrm{v}_n\{\mathrm{EP}\}$. For $\mathrm{v}_n\{\mathrm{EP}\}$, the measurements shown in Figs.~\ref{fig:Fig5} and ~\ref{fig:Fig7} are taken and integrated over $0.5 < \pT <20\,$\gev.  The mean values of the measured $p(\mathrm{v}_n)$ distributions are also shown and marked as $\mathrm{v}_n\{\mathrm{EbyE}\}$. Over the accessible centrality range and for $n=2,3,4$, a systematic pattern is seen with $\mathrm{v}_n\{2\}  > \mathrm{v}_n\{\mathrm{EP}\} \geq \mathrm{v}_n\{\mathrm{EbyE}\} > \mathrm{v}_n\{4\}$. 
The $\mathrm{v}_n\{2\}$ values are the largest (predominantly) due to large contributions from short-range two-particle correlations, which are suppressed in the event-plane $\mathrm{v}_n$ measurements.  The $\mathrm{v}_n$ coefficients measured with the event-plane method are systematically larger than the mean values of the event-by-event measurement of flow harmonics. This difference is naturally attributed to the flow fluctuations, which contribute to $\mathrm{v}_n\{\mathrm{EP}\}$ but are suppressed in $\mathrm{v}_n\{\mathrm{EbyE}\}$.
The flow coefficients measured with the four-particle cumulant method are the smallest, mainly due to the contribution from flow fluctuations, which is negative for $\mathrm{v}_n\{4\}$ and positive for $\mathrm{v}_n$ measured with the event-plane method. In addition, some residual two-particle correlations unrelated to the azimuthal asymmetry in the initial geometry  contribute to $\mathrm{v}_n\{\mathrm{EP}\}$, but are negligibly small in the case of $\mathrm{v}_n\{4\}$.
\begin{figure}[th!]
\begin{center}
\includegraphics[width=80mm]{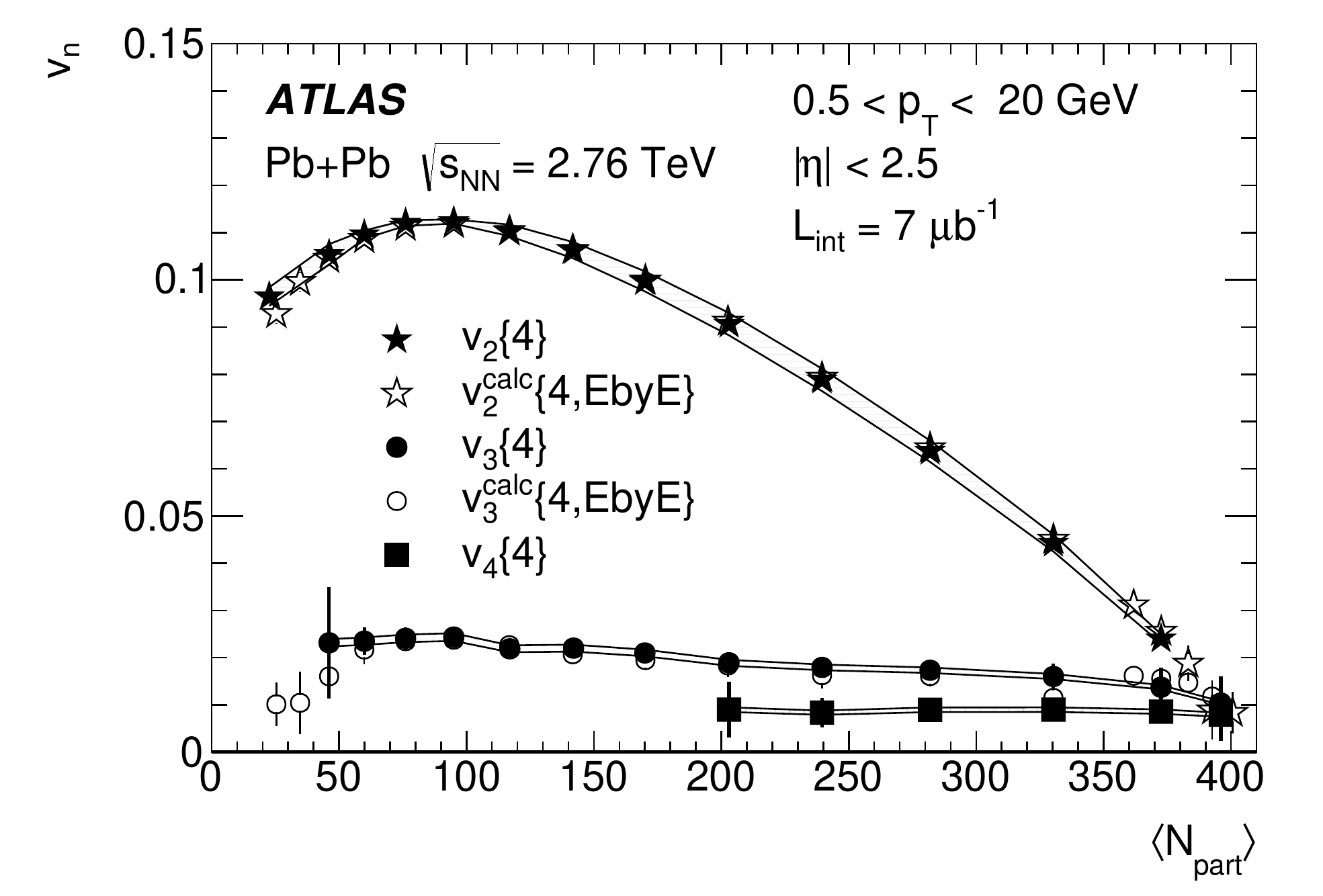}
    \caption{The harmonics $\mathrm{v}_2\{4\}$,  $\mathrm{v}_3\{4\}$ and $\mathrm{v}_4\{4\}$ as a function of $\langle N_{\mathrm{part}}\rangle$. Filled symbols show the results from the cumulant method while open symbols show $\mathrm{v}_{2,3}^{\mathrm{calc}}\{4,\mathrm{EbyE}\}$ calculated from the $p(\mathrm{v}_2)$ and $p(\mathrm{v}_3)$ distributions. Statistical errors are shown as bars and systematic uncertainties as shaded bands. 
}
\label{fig:Fig13}
\end{center}
\end{figure}

The centrality dependence of $\mathrm{v}_n\{4\}$ is shown in Fig.~\ref{fig:Fig13} for $n=2,3$ and 4. The elliptic flow $\mathrm{v}_2\{4\}$ shows a strong centrality dependence, rising with $ N_{\mathrm{part}}$ until reaching a maximum at $N_{\mathrm{part}}\approx 100$, and then decreasing for more central collisions. This strong centrality dependence is not seen for the higher flow harmonics $\mathrm{v}_3\{4\}$ and $\mathrm{v}_4\{4\}$. In addition, the magnitude of the third- and fourth-order flow coefficients is much smaller than the magnitude of the elliptic flow; e.g. for $<N_{\mathrm{part}}>$ of about 300, $\mathrm{v}_n \approx$ 0.05, 0.02, and 0.01 for $\mathrm{v}_2$, $\mathrm{v}_3$ and $\mathrm{v}_4$, respectively. For smaller $N_{\mathrm{part}}$ this difference is even larger, with $\mathrm{v}_2$ reaching more than 0.1 and $\mathrm{v}_3$ and $\mathrm{v}_4$ staying at the same level as at higher $N_{\mathrm{part}}$. Figure~\ref{fig:Fig13} also shows $\mathrm{v}_{2,3}^{\mathrm{calc}}\{4,\mathrm{EbyE}\}$ obtained from the measured $p(\mathrm{v}_2)$ and $p(\mathrm{v}_3)$ distributions. The measured $p(\mathrm{v}_4)$ in Ref. \citep{Atlas3} is truncated at large values of $\mathrm{v}_4$ and therefore is not used here for the comparison.  Good agreement between the two independent measurements is also seen for the third-order flow harmonics. 

\subsection{Fluctuations of flow harmonics}
\label{sec:fluctdep}
Measurements of elliptic flow dynamic fluctuations have attracted much interest, since flow fluctuations can be traced back to fluctuations of the initial collision zone. Experimentally, flow fluctuations are difficult to measure due to unavoidable contamination by non-flow effects. The reported elliptic flow fluctuation measurements from RHIC \citep{phobosfluct1,phobosfluct2,starfluct} are affected by non-flow correlations, despite the attempts made to estimate their contribution.  Interestingly,  RHIC results indicate that flow fluctuations are mostly determined by initial-state geometry fluctuations, which thus seem to be preserved throughout the system evolution.

The relative flow harmonic fluctuations, defined as $\sigma_{\mathrm{v}_n}/\left\langle \mathrm{v}_n \right\rangle$, can be calculated using the width and mean value of the $p(\mathrm{v}_n)$ distributions and compared to the predictions for fluctuations of the initial geometry. The latter can be characterized by the eccentricities, $\epsilon_n$, which can be estimated from the transverse positions $(r,\phi)$ of nucleons participating in the collision:
\begin{equation}
\epsilon_n=\frac{\sqrt{\langle r^n \cos n\phi\rangle^2 + \langle r^n \sin n\phi\rangle^2}}{\langle r^n\rangle}.
\label{eq:ecc}
\end{equation}
\noindent
Such a comparison of $\sigma_{\mathrm{v}_n}/\left\langle \mathrm{v}_n \right\rangle$,  derived from the event-by-event measurement of $\mathrm{v}_n$, to the Glauber model \citep{glauber1} and MC-KLN model \citep{KLN}, which combines the Glauber approach with saturated low-$x$ gluon distribution functions,  is discussed in Ref.~\citep{Atlas3}. In general, none of the considered models of  the relative fluctuations of $\epsilon_n$ gives a consistent description of the relative flow fluctuations over the entire range of collision centralities.  

In this analysis, the measure of relative flow fluctuations, $F(\mathrm{v}_n)$, is defined as:
\begin{equation}
 F(\mathrm{v}_n) = \sqrt{\frac{ \mathrm{v}_n\{2\}^2 - \mathrm{v}_n\{4\}^2}{ \mathrm{v}_n\{2\}^2 + \mathrm{v}_n\{4\}^2}}.
    \label{eq:fluSig}
\end{equation}
\noindent
The above formula provides a valid estimate of $\sigma_{\mathrm{v}_n}/\left\langle \mathrm{v}_n \right\rangle$ under the assumptions that non-flow correlations are absent in $\mathrm{v}_n\{2\}$ and $\mathrm{v}_n\{4\}$, and that flow fluctuations are small  compared to $\langle \mathrm{v}_n \rangle$ ($\sigma_{\mathrm{v}_n} << \langle \mathrm{v}_n \rangle$). The first assumption is obviously not fulfilled by $\mathrm{v}_n\{2\}$, which is strongly contaminated by non-flow correlations. Therefore, $\mathrm{v}_n\{\mathrm{EP}\}$ is used instead of $\mathrm{v}_n\{2\}$, following the approach proposed in Ref.~\citep{Alice3}. The second assumption is not valid for fluctuations of third- and fourth-order flow harmonics, and also for elliptic flow harmonics measured in the most central Pb+Pb collisions. Nevertheless, it is interesting to study this alternative measure of flow fluctuations and to compare it to the same quantity predicted by initial-state models. For head-on nucleus--nucleus collisions, $\langle N_{\mathrm{part}}\rangle \approx 400$, the prediction for eccentricity fluctuations $\sigma_{\epsilon_n}/\left\langle \epsilon_n \right\rangle$ reaches the limit of $\sqrt{4/\pi-1} \approx$ 0.52 for the fluctuations-only scenario when the $\epsilon_n$ distribution  is described by a two-dimensional Gaussian function in the transverse plane~\citep{glissando2}. In Ref.~\citep{Atlas3} it was shown that this limit is indeed reached by $\sigma_{\mathrm{v}_n}/\left\langle \mathrm{v}_n \right\rangle$ for $\mathrm{v}_2$ measured in the most central Pb+Pb collisions and for $\mathrm{v}_3$ and $\mathrm{v}_4$ over the entire centrality range. For the fluctuations-only scenario, the estimate $F(\mathrm{v}_n)$ should be close to one since then $\langle \mathrm{v}_n\{4\}\rangle \approx 0$. Thus, it is interesting to compare this alternative measure of flow fluctuations to the same quantity derived from the initial eccentricity distributions, $F(\epsilon_n)$. It can be seen by comparison with Eq.~(\ref{eq:fluSig}) that the quantity $F$ depends not only on the second moment of the  $\epsilon_n$ distribution (as does $\sigma_{\epsilon_n}/\left\langle \epsilon_n \right\rangle$), but also on the fourth moment and, therefore, can provide a more sensitive test of model assumptions. 
\begin{figure*}[ht!]
\begin{center}
\includegraphics[width=170mm]{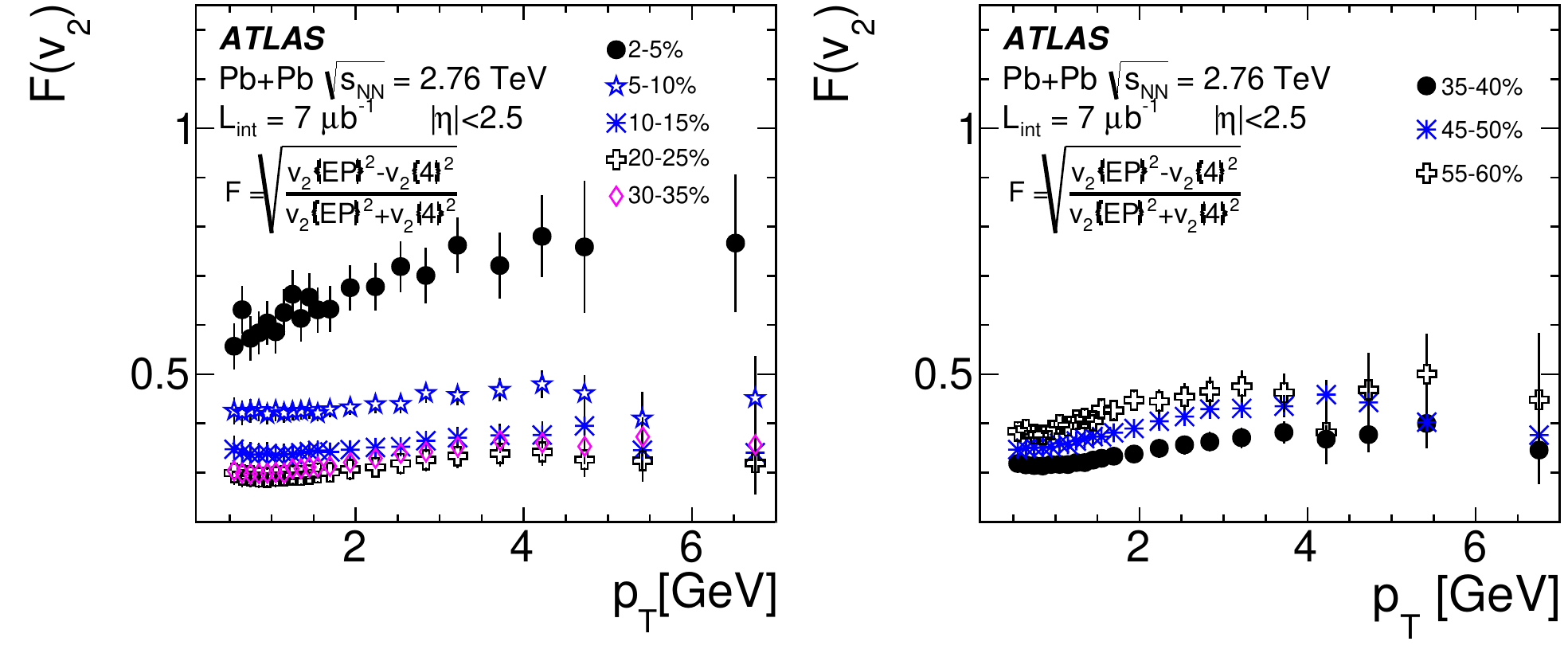}
\caption[flow fluct]{\label{fig:Fig14}
The transverse momentum dependence of the relative elliptic flow fluctuations, as measured by $F$ with $\mathrm{v}_2\{2\}$ replaced by $\mathrm{v}_2\{\mathrm{EP}\}$, for central collisions (left panel) and peripheral collisions (right panel). The error bars denote statistical and systematic uncertainties added in quadrature. 
}
\end{center}
\end{figure*}

Figure~\ref{fig:Fig14} shows the \pT\ dependence of the relative elliptic flow fluctuations calculated for different centrality intervals with Eq.~(\ref{eq:fluSig}), where $\mathrm{v}_2\{2\}$ is replaced by $\mathrm{v}_2\{\mathrm{EP}\}$. For all centrality intervals, except 2--5\%,  the relative elliptic flow fluctuations depend only weakly on \pT\ over the whole \pT\ range, indicating that they are predominantly associated with fluctuations of the initial geometry. A similar \pT\  dependence of the relative elliptic flow fluctuations was  recently reported by the ALICE collaboration~\citep{Alice3}, although the ALICE results for the 0--5\% most central collisions show a much stronger \pT\ dependence than the present measurement for the centrality interval 2--5\%. This discrepancy may be due to different contributions of non-flow effects to $\mathrm{v}_2\{\mathrm{EP}\}$ measured in the two experiments. 

\begin{figure}[ht!]
\begin{center}
\includegraphics[width=78mm]{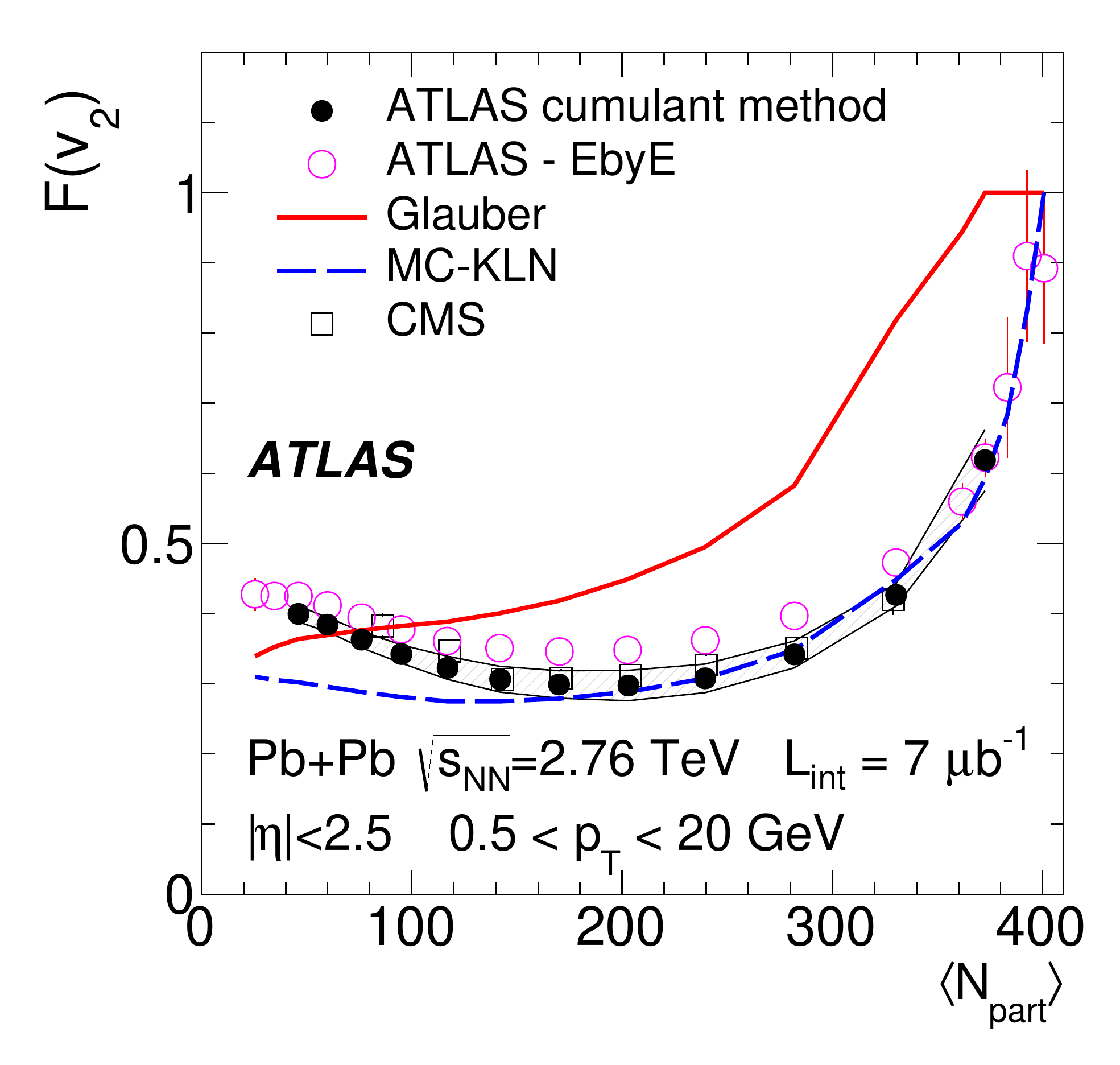}
    \caption{The relative elliptic flow fluctuations,  $F(\mathrm{v}_2)$,   as a function of $\langle N_{\mathrm{part}}\rangle$,  from this analysis (filled circles) with $\mathrm{v}_2\{\mathrm{EP}\}$ substituted for $\mathrm{v}_2\{2\}$. Statistical errors are shown as bars and systematic uncertainties as shaded bands. $F(\mathrm{v}_2)$ obtained from the measured $\mathrm{v}_2$ distribution \citep{Atlas3} is shown as open circles (marked in the legend as ``EbyE") with the error bars denoting statistical and systematic uncertainties added in quadrature. The same quantity calculated from the initial eccentricity distributions obtained from the Glauber~\citep{glauber1} and MC-KLN~\citep{KLN} models is shown by curves. Open squares show the CMS measurement of $F(\mathrm{v}_2)$~\citep{Cms4}.
}
\label{fig:Fig15}
\end{center}
\end{figure}

The quantity $F(\mathrm{v}_n)$ is further investigated as a function of the collision centrality using flow harmonics averaged over \pT\ and $\eta$. The dependence of $F(\mathrm{v}_2)$  on $\langle N_{\mathrm{part}}\rangle$ is shown in Fig.~\ref{fig:Fig15}. Two sets of measurements are shown: $F(\mathrm{v}_2)$ calculated using $\mathrm{v}_2\{4\}$ and $\mathrm{v}_2\{2\}$ obtained with the cumulant method with $\mathrm{v}_2\{\mathrm{EP}\}$ replacing $\mathrm{v}_2\{2\}$, and using $\mathrm{v}_2^{\mathrm{calc}}\{4,\mathrm{EbyE}\}$ and $\mathrm{v}_2^{\mathrm{calc}}\{2,\mathrm{EbyE}\}$ obtained from the measured $p(\mathrm{v}_2)$ distribution \citep{Atlas3}.  
The two measurements show similar centrality dependence, but the estimate based on the cumulant method is systematically smaller (by up to about 15\%) than that calculated from $p(\mathrm{v}_2)$. 
\begin{figure}[ht!]
\begin{center}
\includegraphics[width=78mm]{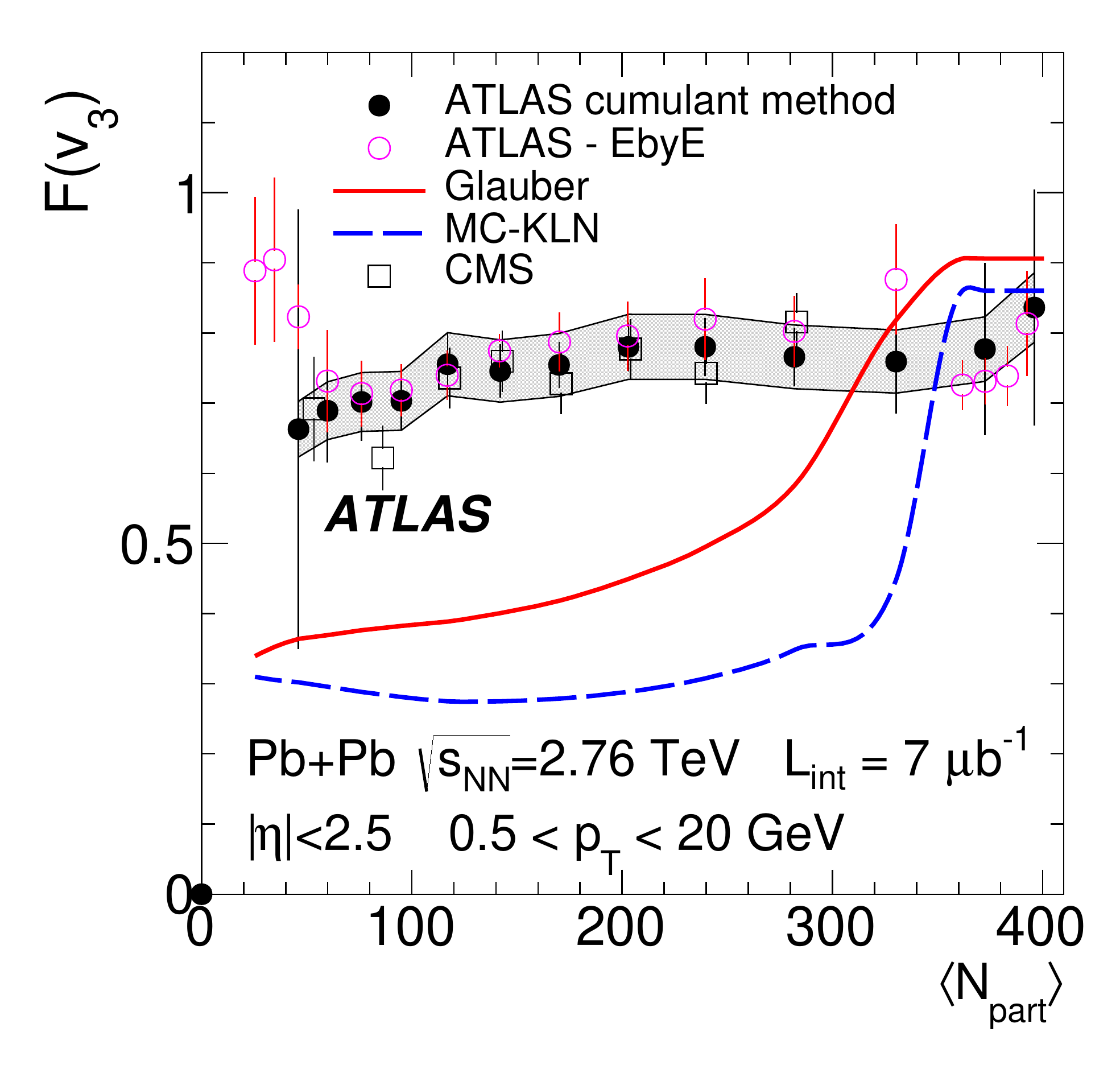}
\includegraphics[width=78mm]{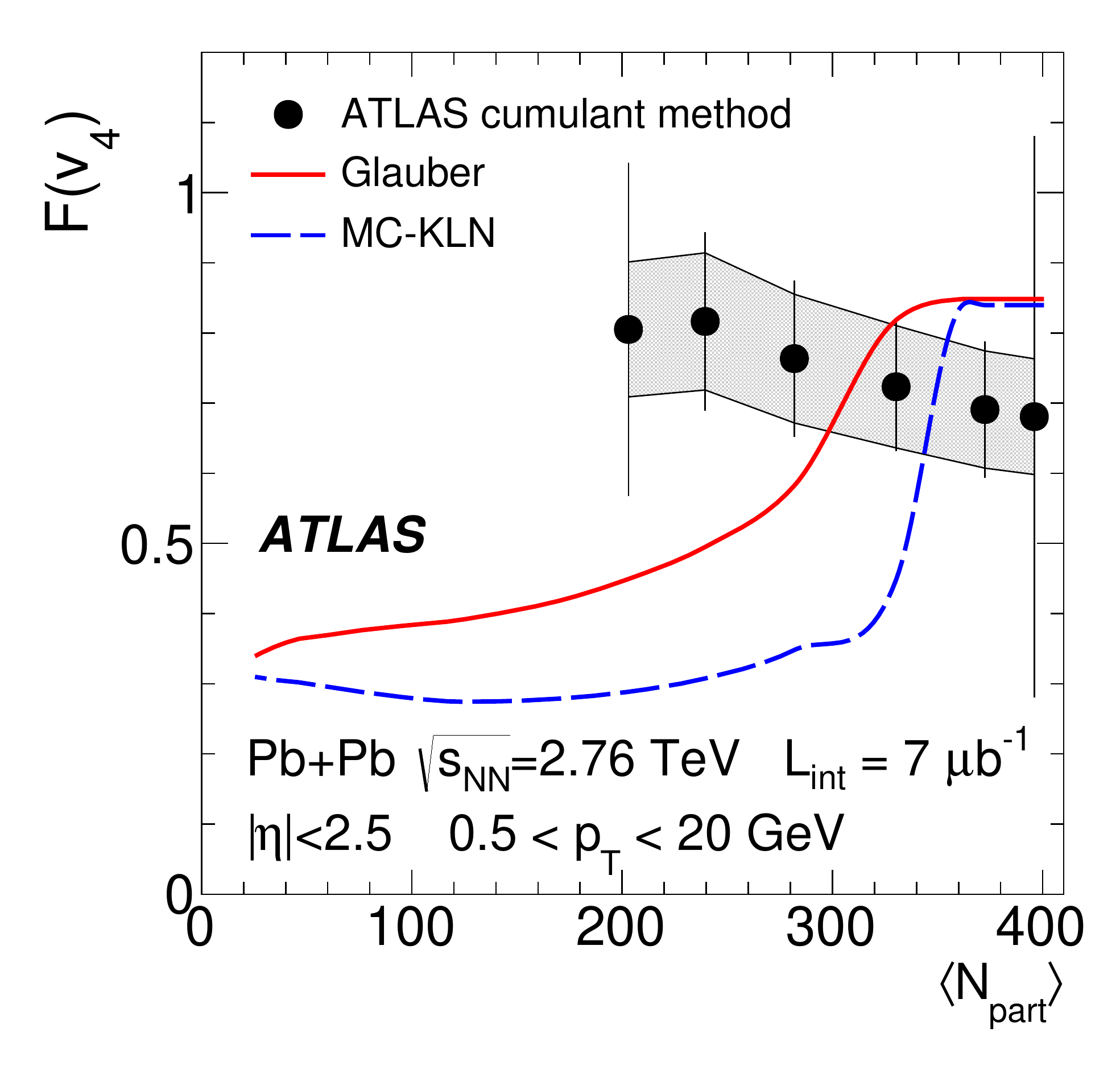}
    \caption{The relative fluctuations,  $F(\mathrm{v}_n)$  for $n=3$ (top)  and  $n=4$ (bottom) as a function of $\langle N_{\mathrm{part}}\rangle$, from the ATLAS cumulant method (filled circles) with $\mathrm{v}_n\{\mathrm{EP}\}$ substituted for $\mathrm{v}_n\{2\}$. Statistical errors are shown as bars and systematic uncertainties as shaded bands. The same quantity calculated from the measured $\mathrm{v}_3$ distributions \citep{Atlas3} is shown as open circles in the top plot, with error bars denoting statistical and systematic uncertainties added in quadrature. The top plot also shows $F(\mathrm{v}_3)$ as measured by CMS \citep{Cms4} (open squares). $F$ values calculated from the eccentricity distributions obtained from the Glauber~\citep{glauber1} and MC-KLN~\citep{KLN} models  are shown by curves.  
}
\label{fig:Fig16}
\end{center}
\end{figure}

$F(\mathrm{v}_2)$ can also be compared to 
$\sigma_{\mathrm{v}_2}/\left\langle \mathrm{v}_2\right\rangle$ determined from the $p(\mathrm{v}_2)$ distribution. It was shown in Ref.~\citep{Atlas3} that the two measures of elliptic flow fluctuations agree for the most peripheral collisions. For semi-central collisions,  
$\sigma_{\mathrm{v}_2}/\left\langle \mathrm{v}_2 \right\rangle$ is systematically larger than $F(\mathrm{v}_2)$. A significant discrepancy was shown for the most central collisions, where $\sigma_{\mathrm{v}_2}/\left\langle \mathrm{v}_2 \right\rangle$ levels off at 0.52, 
while $F(\mathrm{v}_2)$ rises continuously, reaching much higher values.  

  $F(\mathrm{v}_2)$ shows a strong centrality dependence. Starting with the most peripheral collisions, it decreases with $\langle N_{\mathrm{part}}\rangle$, reaching a minimum at $\langle N_{\mathrm{part}}\rangle \approx 200$ and then rises steeply with centrality up to about 1.0 for the most central collisions. A comparison to $F(\epsilon_2)$ calculated from the eccentricity distributions predicted by the Glauber and MC-KLN models is shown in Fig.~\ref{fig:Fig15}. One can see that the MC-KLN model describes the measurements for Pb+Pb collisions with $\langle N_{\mathrm{part}}\rangle$ above 150 reasonably well, while the Glauber model significantly over-predicts the measured $F(\mathrm{v}_2)$ across the entire centrality range. Figure~\ref{fig:Fig15} also shows that our results are consistent with the CMS estimate of $F(\mathrm{v}_2)$~\citep{Cms4}.

The study of $F(\mathrm{v}_n)$ is also performed  for higher-order flow harmonics, $n=3,4$. Figure~\ref{fig:Fig16} shows $F(\mathrm{v}_3)$ (top plot) and $F(\mathrm{v}_4)$ (bottom plot) obtained using the cumulant method with $\mathrm{v}_3\{2\}$ and $\mathrm{v}_4\{2\}$ replaced by $\mathrm{v}_3\{\mathrm{EP}\}$ and $\mathrm{v}_4\{\mathrm{EP}\}$, respectively. Large relative fluctuations of the third- and fourth-order harmonics, of the order of 0.7--0.8, are measured over the whole accessible centrality range, with a relatively weak centrality dependence. The results are consistent with the CMS measurement of $F(\mathrm{v}_3)$ \citep{Cms4} as well as with  $F$ calculated using $\mathrm{v}_3^{\mathrm{calc}}\{4,\mathrm{EbyE}\}$ and $\mathrm{v}_3^{\mathrm{calc}}\{2,\mathrm{EbyE}\}$ \citep{Atlas3}. $F(\epsilon_3)$ and $F(\epsilon_4)$ obtained from the eccentricity distributions predicted by the Glauber and MC-KLN models are also shown. One can see that none of the models gives a consistent description of $F(\mathrm{v}_3)$ and $F(\mathrm{v}_4)$.  
  
\section{Summary}
\label{sec:summary}
A measurement of flow harmonics of charged particles in Pb+Pb collisions at $\sqrt{s_{\mathrm{NN}}}$ = 2.76~\TeV\ from the ATLAS experiment at the LHC is presented using a dataset of approximately 7 $\mu$b$^{-1}$ collected in 2010. The analysis is based on the cumulant expansion of multi-particle azimuthal correlations, which suppresses correlations not related to the initial-state geometry. Another advantage of the cumulant method is that it provides several different measurements  of the same harmonic, $\mathrm{v}_n$, allowing for estimation of the non-flow contributions and for consistency checks. The need for  huge computing power in calculating multi-particle correlations is overcome by using the generating function formalism.

Flow coefficients $\mathrm{v}_n$ for $n=2,3,4$ were obtained using two- and four-particle cumulants. In addition, for the elliptic flow ($n=2$), the analysis is for the first time extended to the six- and eight-particle cumulants. The transverse momentum, pseudorapidity and centrality dependence of flow harmonics is presented. An attempt is also made to estimate the flow harmonic fluctuations using the measured $\mathrm{v}_n\{4\}$ and $\mathrm{v}_n$ obtained with the event-plane method. 
 
The transverse momentum dependence of $\mathrm{v}_2\{2\}$ shows significant non-flow contributions. This contribution is reduced in $\mathrm{v}_2\{\mathrm{EP}\}$. The elliptic flow obtained with the four-particle cumulants provides a measure of $\mathrm{v}_2$ with non-flow correlations strongly suppressed. Using six- and eight-particle cumulants gives results consistent, within the errors, with those obtained with four-particle cumulants, indicating that four-particle cumulants efficiently suppress non-flow correlations. Similar conclusions can be drawn from the study of the $\eta$-dependence of the $\pT$-integrated $\mathrm{v}_2$ as well as from the centrality dependence of $\mathrm{v}_2$ averaged over \pT\ and $\eta$. As for $\mathrm{v}_2$, the higher-order flow harmonics determined using four-particle cumulants are significantly reduced compared to the measurement involving two-particle cumulants. 

The flow harmonics, $\mathrm{v}_n\{4\}$, determined from the four-particle cumulants increase sharply with \pT\, reaching a maximum at 2--3~\gev. At higher transverse momenta, $\mathrm{v}_2\{4\}$ decreases  and beyond \pT\ of about 7~\gev, it plateaus at the level of about 0.04 up to the highest accessible transverse momenta. The higher-order harmonics also decrease above 3~\gev\ and reach a value of about 0.03 when integrated over \pT\ from 4 \gev\ to 20 \gev. The four-particle harmonics, $\mathrm{v}_n\{4\}$, are found to depend weakly on the pseudorapidity over the range $|\eta|<2.5$.

The centrality dependence of the \pT- and $\eta$-integrated $\mathrm{v}_n\{4\}$ reveals a clear distinction between $\mathrm{v}_2$ and the higher-order harmonics: $\mathrm{v}_2$ strongly depends on the collision centrality, reflecting its sensitivity to the varying shape of the initial collision geometry, while $\mathrm{v}_3$ and $\mathrm{v}_4$
show only a weak centrality dependence, predominantly attributed to  geometry fluctuations. Over the studied centrality range, except for the most central collisions, the measured $\mathrm{v}_3$ and $\mathrm{v}_4$ are much smaller than $\mathrm{v}_2$. For the most central collisions, a similar magnitude is measured for $\mathrm{v}_2$, $\mathrm{v}_3$  and $\mathrm{v}_4$ because all are dominated by geometry fluctuations. The measured $\mathrm{v}_2\{2k\}$ for $k=2,3,4$ and $\mathrm{v}_3\{4\}$ are found to agree with the same coefficients calculated from the moments of the measured $p(\mathrm{v}_2)$ and $p(\mathrm{v}_3)$ distributions.

The relative flow harmonic fluctuations, $F(\mathrm{v}_n)$, defined in Eq.~(\ref{eq:fluSig}),  are estimated using $\mathrm{v}_n\{\mathrm{EP}\}$ and $\mathrm{v}_n\{4\}$.  For the elliptic flow harmonic, a strong centrality dependence is observed, following a trend similar to that exhibited by $F$ as estimated from the $p(\mathrm{v}_2)$ distribution. In contrast, $F(\mathrm{v}_3)$ and $F(\mathrm{v}_4)$ show a weak centrality dependence. The large magnitudes of $F$ obtained for third- and fourth-order harmonics, and also for the elliptic flow harmonic measured in the most central collisions, indicate the dominant role of initial-state fluctuations. The comparison to the same quantity derived from the initial-state eccentricity distributions, modelled by the Glauber and MC-KLN models, shows that none of these models can  describe the flow harmonic fluctuations well, particularly for higher-order flow harmonics. Therefore, the measurements presented in this paper provide valuable constraints on models of initial spatial anisotropy  and subsequent hydrodynamic evolution of systems produced in ion--ion collisions with nucleon--nucleon centre-of-mass energies at the TeV energy scale.

\section{Acknowledgements}
\label{sec:ack}

We thank CERN for the very successful operation of the LHC, as well as the
support staff from our institutions without whom ATLAS could not be
operated efficiently.

We acknowledge the support of ANPCyT, Argentina; YerPhI, Armenia; ARC,
Australia; BMWFW and FWF, Austria; ANAS, Azerbaijan; SSTC, Belarus; CNPq and FAPESP,
Brazil; NSERC, NRC and CFI, Canada; CERN; CONICYT, Chile; CAS, MOST and NSFC,
China; COLCIENCIAS, Colombia; MSMT CR, MPO CR and VSC CR, Czech Republic;
DNRF, DNSRC and Lundbeck Foundation, Denmark; EPLANET, ERC and NSRF, European Union;
IN2P3-CNRS, CEA-DSM/IRFU, France; GNSF, Georgia; BMBF, DFG, HGF, MPG and AvH
Foundation, Germany; GSRT and NSRF, Greece; ISF, MINERVA, GIF, I-CORE and Benoziyo Center,
Israel; INFN, Italy; MEXT and JSPS, Japan; CNRST, Morocco; FOM and NWO,
Netherlands; BRF and RCN, Norway; MNiSW and NCN, Poland; GRICES and FCT, Portugal; MNE/IFA, Romania; MES of Russia and ROSATOM, Russian Federation; JINR; MSTD,
Serbia; MSSR, Slovakia; ARRS and MIZ\v{S}, Slovenia; DST/NRF, South Africa;
MINECO, Spain; SRC and Wallenberg Foundation, Sweden; SER, SNSF and Cantons of
Bern and Geneva, Switzerland; NSC, Taiwan; TAEK, Turkey; STFC, the Royal
Society and Leverhulme Trust, United Kingdom; DOE and NSF, United States of
America.

The crucial computing support from all WLCG partners is acknowledged
gratefully, in particular from CERN and the ATLAS Tier-1 facilities at
TRIUMF (Canada), NDGF (Denmark, Norway, Sweden), CC-IN2P3 (France),
KIT/GridKA (Germany), INFN-CNAF (Italy), NL-T1 (Netherlands), PIC (Spain),
ASGC (Taiwan), RAL (UK) and BNL (USA) and in the Tier-2 facilities
worldwide.

\vspace{-6mm}
\bibliographystyle{atlasBibStyleWoTitle}
\bibliography{CumulantPaperPbPb}

\onecolumn
\clearpage
\begin{flushleft}
{\Large The ATLAS Collaboration}

\bigskip

G.~Aad$^{\rm 84}$,
B.~Abbott$^{\rm 112}$,
J.~Abdallah$^{\rm 152}$,
S.~Abdel~Khalek$^{\rm 116}$,
O.~Abdinov$^{\rm 11}$,
R.~Aben$^{\rm 106}$,
B.~Abi$^{\rm 113}$,
M.~Abolins$^{\rm 89}$,
O.S.~AbouZeid$^{\rm 159}$,
H.~Abramowicz$^{\rm 154}$
% ...
\bigskip
\\
% ...
$^{11}$ Institute of Physics, Azerbaijan Academy of Sciences, Baku, Azerbaijan\\
% ...
$^{84}$ CPPM, Aix-Marseille Universit{\'e} and CNRS/IN2P3, Marseille, France\\
% ...
$^{89}$ Department of Physics and Astronomy, Michigan State University, East Lansing MI, United States of America\\
% ...
$^{106}$ Nikhef National Institute for Subatomic Physics and University of Amsterdam, Amsterdam, Netherlands\\
% ...
$^{112}$ Homer L. Dodge Department of Physics and Astronomy, University of Oklahoma, Norman OK, United States of America\\
$^{113}$ Department of Physics, Oklahoma State University, Stillwater OK, United States of America\\
% ...
$^{116}$ LAL, Universit{\'e} Paris-Sud and CNRS/IN2P3, Orsay, France\\
% ...
$^{152}$ Institute of Physics, Academia Sinica, Taipei, Taiwan\\
% ...
$^{154}$ Raymond and Beverly Sackler School of Physics and Astronomy, Tel Aviv University, Tel Aviv, Israel\\
% ...
$^{159}$ Department of Physics, University of Toronto, Toronto ON, Canada\\
% ...
\end{flushleft}

\end{document}